\documentclass[aps,prb,amsmath,twocolumn,superscriptaddress]{revtex4-2}

\usepackage{amsmath}
\usepackage{amssymb}
\usepackage{hyperref}
\usepackage{tikz}

\usepackage{graphicx}
\usepackage{bm} 
\usepackage{url}
\usepackage{verbatim}   

\usepackage[normalem]{ulem}
\usepackage{color}

\usepackage{amsmath}

\usepackage[percent]{overpic}
\usepackage[export]{adjustbox}
\usepackage{stackengine}
\usepackage{appendix}

\usepackage{placeins} 
\usepackage{flafter}  

\begin{document}

\title{Transient ordering in the Gross-Pitaevskii lattice after an energy quench within a non-ordered phase}

\author{Andrei E. Tarkhov}
\email{atarkhov@bwh.harvard.edu}
\affiliation{Moscow Institute of Physics and Technology, Institutskiy per.
9, Dolgoprudny, Moscow region 141700, Russia}
\affiliation{Division of Genetics, Department of Medicine, Brigham and Women’s Hospital and Harvard Medical School, Boston, MA 02115, USA}
\affiliation{Skolkovo Institute of Science and Technology, Bolshoy Boulevard 30, bld. 1, 121205 Moscow, Russia}

\author{A.V. Rozhkov}%
\affiliation{Moscow Institute of Physics and Technology, Institutskiy per.
9, Dolgoprudny, Moscow region 141700, Russia}
\affiliation{Institute for Theoretical and Applied Electrodynamics, Russian
Academy of Sciences, Moscow 125412, Russia}

\author{Boris V. Fine}
\email{boris.fine@uni-leipzig.de}
\affiliation{Moscow Institute of Physics and Technology, Institutskiy per.
9, Dolgoprudny, Moscow region 141700, Russia}
\affiliation{Skolkovo Institute of Science and Technology, Bolshoy Boulevard 30, bld. 1, 121205 Moscow, Russia}
\affiliation{Institute for Theoretical Physics, University of Leipzig,
Br{\"{u}}derstr. 16, 04103 Leipzig, Germany}%

\date{\today}

\begin{abstract}
We numerically investigate heating-and-cooling quenches taking place entirely in the non-ordered phase of the discrete Gross-Pitaevskii equation on a three-dimensional cubic lattice. In equilibrium, this system exhibits a $U(1)$-ordering phase transition at an energy density which is significantly lower than the minimum one during the quench. Yet, we observe that the post-quench relaxation is accompanied by a transient $U(1)$ ordering, namely, the correlation length of $U(1)$ fluctuations significantly exceeds its equilibrium pre-quench value. The longer and the stronger the heating stage of the quench, the stronger the $U(1)$ transient ordering.  We identify the origin of this ordering with the emergence of a small group of slowly relaxing lattice sites accumulating a large fraction of the total energy of the system.  Our findings suggest that the transient ordering may be a robust feature of a broad class of physical systems. This premise is
consistent with the growing experimental evidence of the transient $U(1)$ order in rather dissimilar settings.
\end{abstract}

\maketitle

\textit{Introduction. ---} 
Response of an interacting many-body system  to a sudden change of external conditions, a quench, has recently become a subject of intense experimental and
theoretical research~\cite{ligh_ind_supercond_cuprat2011exper,
non_eq_supercond2015exper, mitrano2016possible, cantaluppi2018pressure,
budden2021evidence, zhou2006long, kogar2020light,
zong2018ultrafast,zong2019dynamical, zong2019evidence,
yusupov_nat_phys2010exp, order_melt2020prb_essler,
supercond_quench_theor2018lemonik, pump_domain_motion2020theor_millis,
dolgirev2020amplitude, quench_dolgirev2020theory}. 
Thermalization after a quench may take a very long
time~\cite{kogar2020light, quench_dolgirev2020theory}
and exhibit a rich variety of transient
regimes~\cite{ligh_ind_supercond_cuprat2011exper, non_eq_supercond2015exper,
mitrano2016possible, cantaluppi2018pressure, budden2021evidence,
zhou2006long, kogar2020light, ni1997transient, ni1998metastable,
gilhoj1995overshooting, transient_order_num2019latt_gas,
quench_transient2020theor_millis, dolgirev2020amplitude,
quench_dolgirev2020theory}.
It can also be accompanied by the spontaneous formation of inhomogeneous
structures~\cite{zong2019evidence,kogar2020light},
the latter being regularly discussed in numerous papers dedicated, among
other topics, to superconducting, charge-ordering, and magnetic
transitions. There is also a mounting experimental
evidence~\cite{ligh_ind_supercond_cuprat2011exper,
non_eq_supercond2015exper, mitrano2016possible, cantaluppi2018pressure,
budden2021evidence, zhou2006long, kogar2020light}
that a large variety of many-body systems may exhibit non-trivial transient
ordering in response to a quench. The proposed interpretations of the
non-equilibrium transient order
revival/enhancement~\cite{quench_transient2020theor_millis,kogar2020light,
ni1997transient}
are often based on the notion of multiple orders competing against each
other both thermodynamically and kinetically.
%

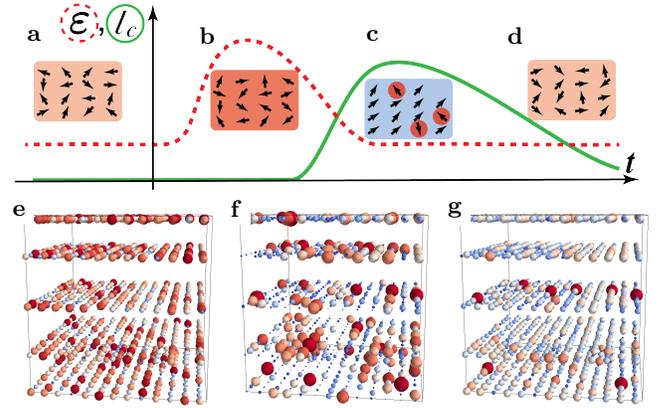
\begin{figure}[h!]
    \begin{tikzpicture}
    \node[inner sep=0pt] (duck) at (0,0)
    {\includegraphics[width=0.96\columnwidth]{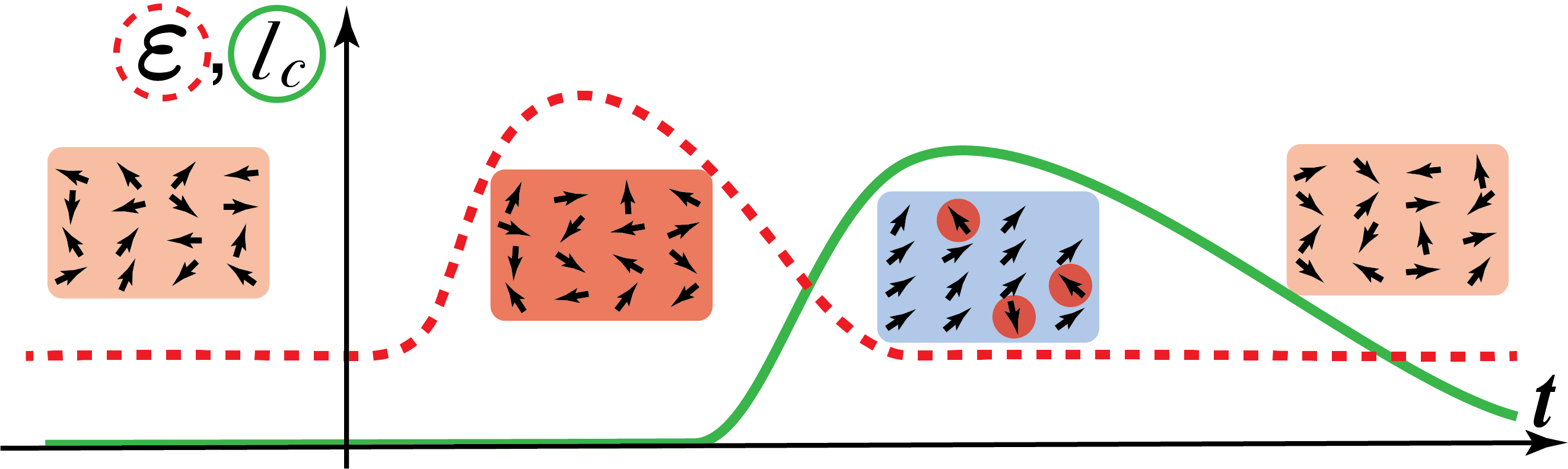}};
    \node[align=center,fill=white] at (-3.9, 0.8) {\textbf{a}};
    \node[align=center,fill=white] at (-1.6, 0.8) {\textbf{b}};
    \node[align=center,fill=white] at (0.6, 0.8) {\textbf{c}};
    \node[align=center,fill=white] at (2.5, 0.8) {\textbf{d}};
    \end{tikzpicture}
    
    \begin{tikzpicture}
    \node[inner sep=0pt] (duck) at (0,0)
    {\includegraphics[width=0.3\columnwidth]{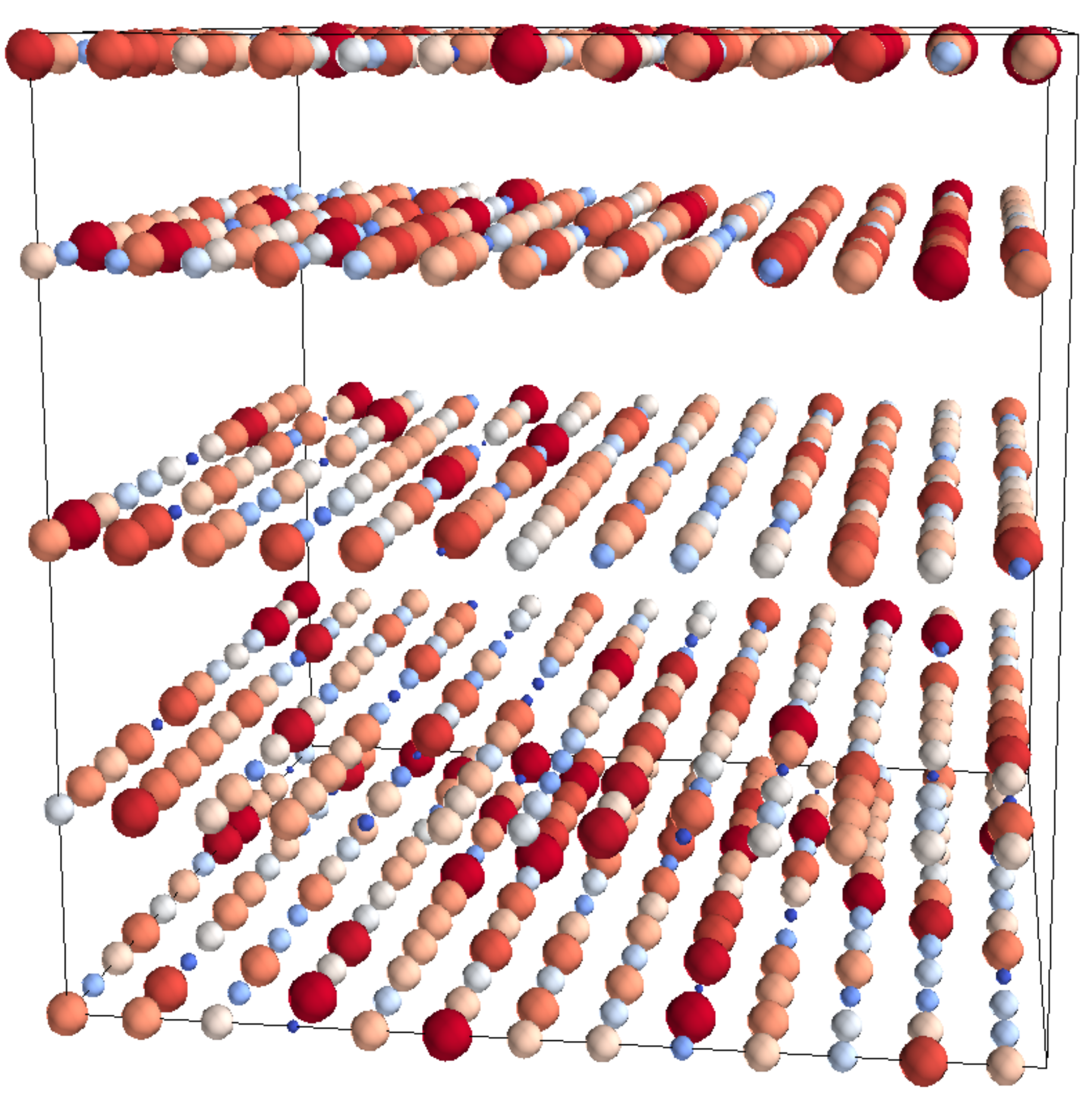}};
    \node[align=center,fill=white] at (-1.3, 1.3) {\textbf{e}};
    \end{tikzpicture}
    \begin{tikzpicture}
    \node[inner sep=0pt] (duck) at (0,0)
    {\includegraphics[width=0.3\columnwidth]{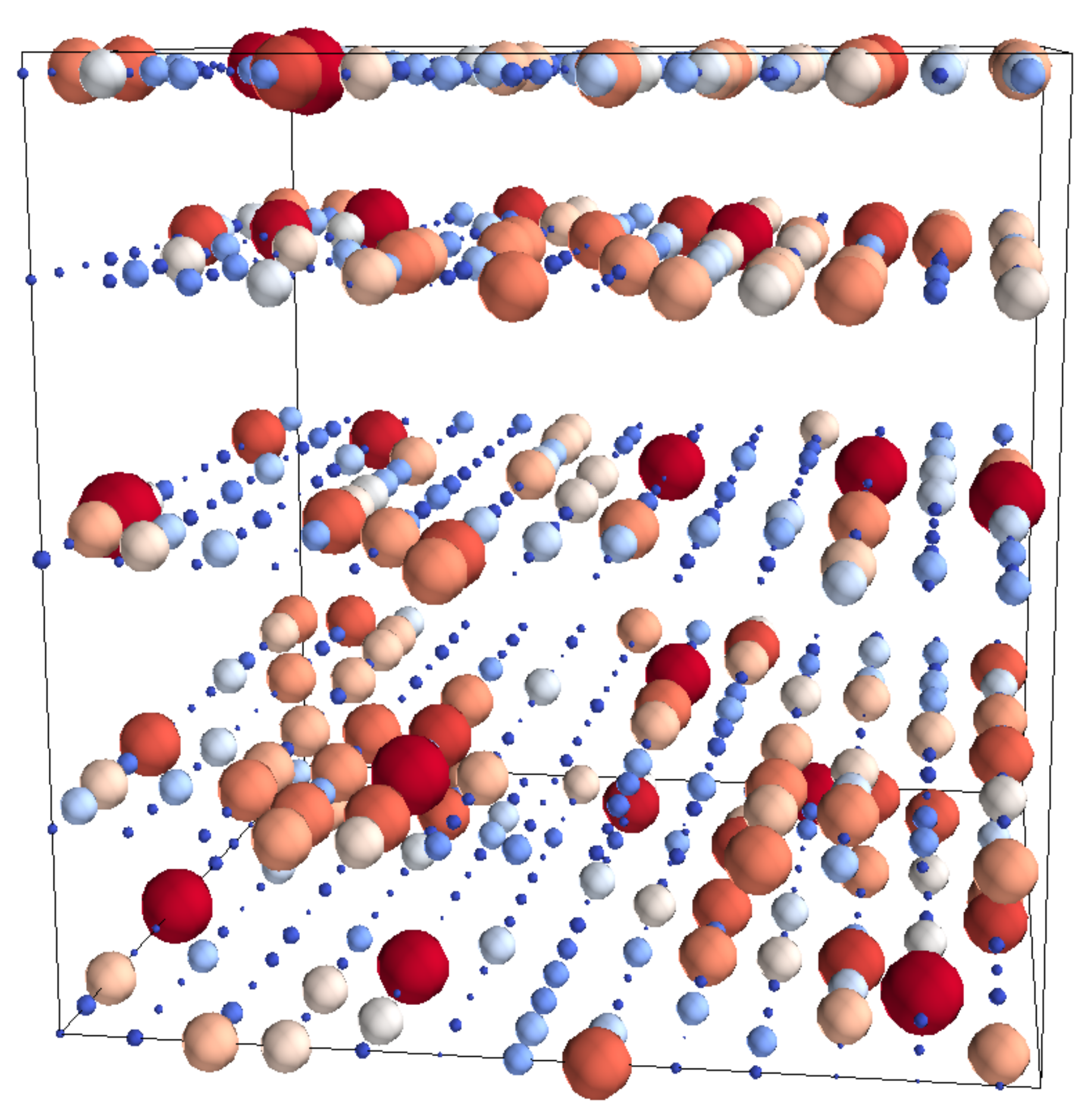}};
    \node[align=center,fill=white] at (-1.3, 1.3) {\textbf{f}};
    \end{tikzpicture}
    \begin{tikzpicture}
    \node[inner sep=0pt] (duck) at (0,0)
    {\includegraphics[width=0.3\columnwidth]{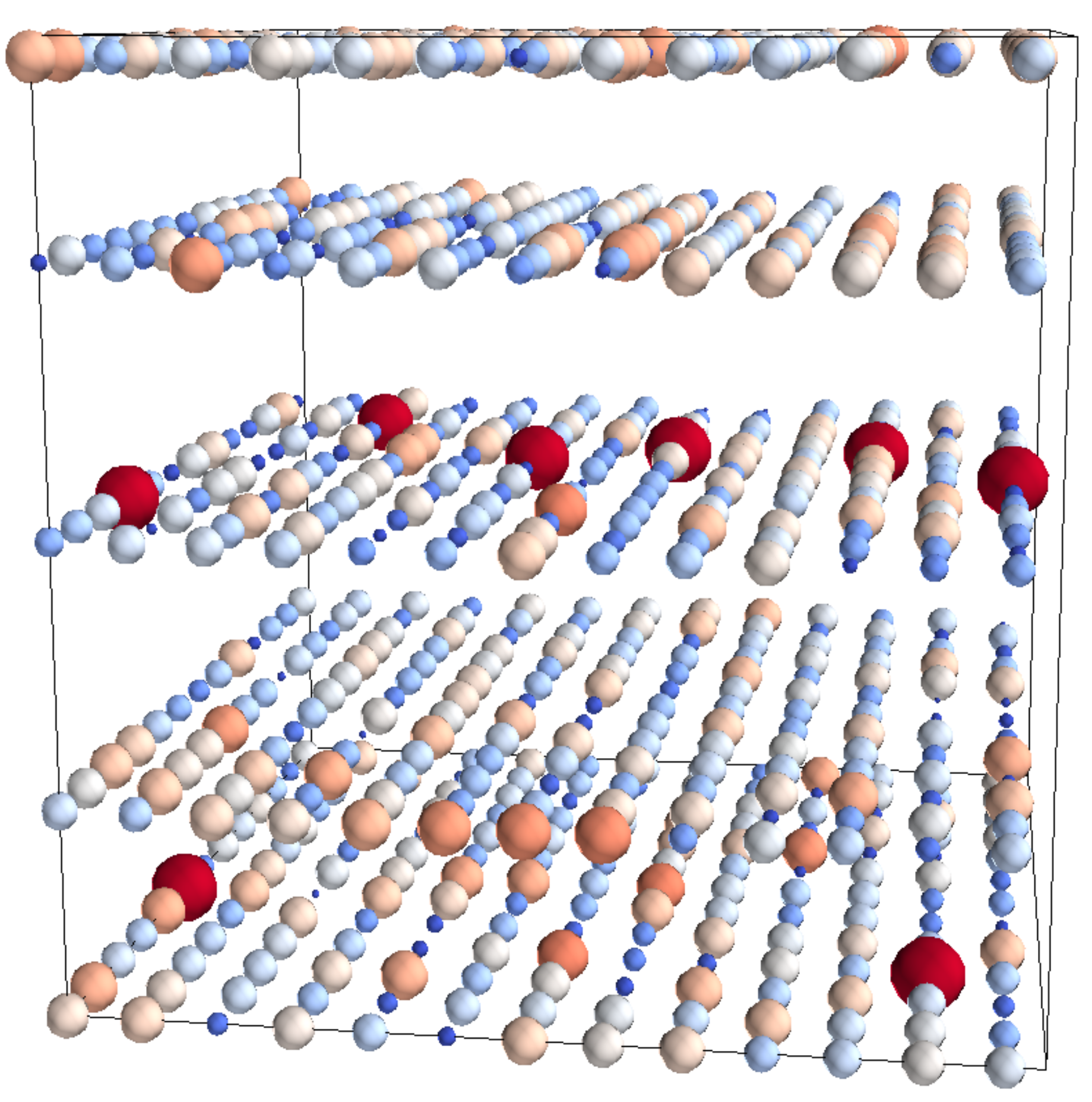}};
    \node[align=center,fill=white] at (-1.3, 1.3) {\textbf{g}};
    \end{tikzpicture}
    
\caption{
\label{fig::sketch_small}
Sketch of the quench procedure. The quench starts at 
$t=0$. The system energy (dashed red line) initially grows, then it reverts to the
pre-quench value. Due to a high pre-quench energy density, the correlation length (green line) is small before the quench, then grows rapidly due to
the quench, relaxing back to the initial value after a long relaxation period. The
cartoons (a-d) depict internal system states at different stages of the quench.
The color represents local energy density, with dark blue being the lowest,
and bright red being the highest, the arrows correspond to the DGPE complex
variables $\psi_j$. The equilibrium state, with moderately high energy density and vanishing
correlation length, is shown in~(a,d,e). Hot mid-quench state is in~(b,f).
Transient ordering is visualized in~(c,g): the arrow directions are
ordered, the energy density is inhomogeneous, with lumps of high energy
shown by red color, the rest of the system being significantly cooler.
(e-g)~Visualization of the DGPE simulation. Local values of 
$n_j$
are shown by volume and color of the balls: blue (red) represents ``cold" 
(``hot") sites.
}
\end{figure}

In this paper, we show that, quite surprisingly, a very simple theoretical model with a single
ordered phase is sufficient for observing the transient ordering. We numerically simulate the
discrete Gross-Pitaevskii equation (DGPE) on a three-dimensional (3D)
lattice, which exhibits an equilibrium phase transition below a certain
temperature. In
Fig.~\ref{fig::sketch_small},
we sketch the transient ordering of the system in the process of quenching. During the quench, the system is subjected to fast heating
followed by fast cooling which ultimately brings the energy back to the
pre-quench value.
It is worth noting that before, during, and after the quench the energy remains within the high-temperature non-ordered phase on the phase
diagram. Yet, we observe that the transient order, absent in
equilibrium, emerges
during the quench, and persists long after that. We argue that the
mechanism behind the observed transient ordering involves the emergence of
a small number of lattice sites with an atypically high concentration of
energy. These sites are the 3D counterparts of the so-called discrete
breathers that are shown to slow down thermalization in 1D DGPE
chains~\cite{dauxois1993energy, mackay1994proof, flach1998discrete,
	campbell2004localizing,rumpf2004simple, ivanchenko2004discrete,
	flach2008discrete,  rumpf2008transition, rumpf2009stable,
	kati2020equilibrium}.
Their concentration may be as little as several per cent, yet, they pull a
significant fraction of total energy from the rest of the system, thereby
temporarily cooling the
latter, which in
turn leads to the detected transient order. Below, we present the detailed description of our
simulations, substantiate our conclusions about the  transient ordering
mechanism and, finally, discuss the implications of our findings.

\textit{The model. ---}
\begin{figure}[h!]
\centering
\includegraphics[width=1\columnwidth]{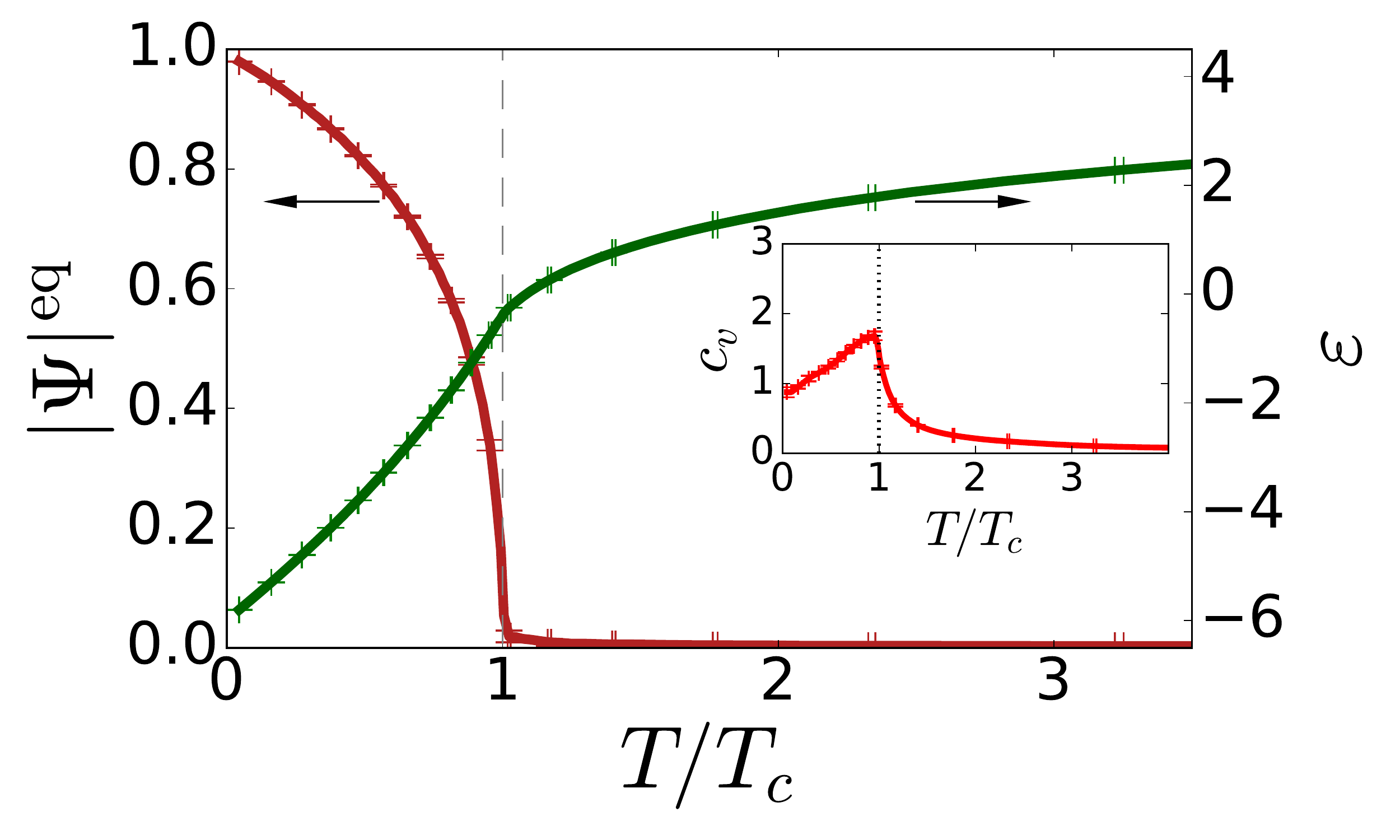}
\caption{
\label{fig::equil_props}
Temperature dependence of the energy density, $\varepsilon$, order parameter, $|\Psi|^{\rm eq}$, and the specific heat, $c_v$ (inset).
}
\end{figure}
The DGPE system on a 3D cubic lattice is a classical dynamical system
describing evolution of complex variables
$\psi_j (t)$
by the following equations
\begin{eqnarray} 
\label{eq::effect_dgpe}
i\frac{d\psi_{j}}{dt}
=
-\sum_{m \in {\rm NN}(j)}
	\psi_{m}+g\left|\psi_{j}\right|^{2}\psi_{j},
\end{eqnarray} 
where $g$ is the interaction parameter, indices $j$ and $m$ label sites of the underlying 3D lattice, and notation
${\rm NN} (j)$
refers to all sites that are nearest-neighbors to site $j$. The DGPE
conserves total energy 
$E = E^{\rm (kin)} + E^{\rm (pot)}$,
where the kinetic energy is $E^{\rm (kin)} = - \sum_j \sum_{m \in {\rm NN}(j)} \psi_m^* \psi_j^{\vphantom{\dagger}},$ and the potential energy is $
E^{\rm (pot)} = \frac{g}{2} \sum_j |\psi_j|^4.$
The norm (also called the ``total number of particles")
$N = \sum_{j} n_j = \sum_{j} \left|\psi_{j}\right|^{2}$
is another integral of motion associated with the invariance
of
Eq.~(\ref{eq::effect_dgpe})
relative to the global ``gauge transformation''
$\psi_j \rightarrow e^{i \alpha} \psi_j$. In this work, we fix $g=10$ and $N=V$,
where $V$ is the total number of sites ($V=50\times50\times50$). The energy density,
$\varepsilon=E/V - g/2$,
is the only parameter that is being changed in the process of quenching.

\textit{Equilibrium state of DGPE system. ---}
For sufficiently large lattice sizes and generic initial conditions
the DGPE dynamics is chaotic, and exhibits ergodization that has been checked by various numerical ergodicity
tests~\cite{tarkhov2017extracting, tarkhov2018estimating, mithun2018weakly,cherny2019non, tarkhov2020ergodization}.
The averaged dynamics may be characterized in terms of entropy and
temperature~\cite{supplemental}
within microcanonical thermodynamic formalism. In equilibrium the
microcanonical temperature $T$ and $\varepsilon$ are connected by
a monotonically growing invertible function
$\varepsilon = \varepsilon^{\rm eq} (T)$.
This function, evaluated numerically and shown in
Fig.~\ref{fig::equil_props}
fora
$g=10$,
 displays an inflection point accompanied by a cusp of the specific heat, $c_v \equiv d\varepsilon(T)/dT$, at
$T_c \approx 4.25$. It is associated with the transition into a low-temperature ordered state,
characterized by the
$U(1)$
order parameter 
$\Psi (t) = |\Psi (t) | e^{i \phi(t)}  = \frac{1}{V} \sum_j \psi_j (t).$
Equilibrium order parameter
$|\Psi| = |\Psi|^{\rm eq} (T)$
is a decreasing function of $T$ for
$T< T_{\rm c}$,
vanishing above
$T_{\rm c}$. 

We also define the correlation length $l_c$ of the $U(1)$ order  as the characteristic length of the decay of the correlation function $\langle \psi^*_m \psi_j \rangle$. The specific definition is $l_c \equiv 1/ \Delta k$, where $\Delta k$ is the half-width at the half-maximum for the Fourier-spectrum intensity $|\psi ({\bf k})|^2$
 around the peak at the wave vector ${\bf k} = {\bf 0}$ \cite{supplemental}. This peak has a finite width above $T_c$, implying that $l_c$ has a finite value, which increases as $T$ decreases towards $T_c$. 

\begin{figure*}[t]
    \centering
    \begin{tikzpicture}
    \node[inner sep=0pt] (duck) at (0,0)
    {\includegraphics[width=0.9\columnwidth]{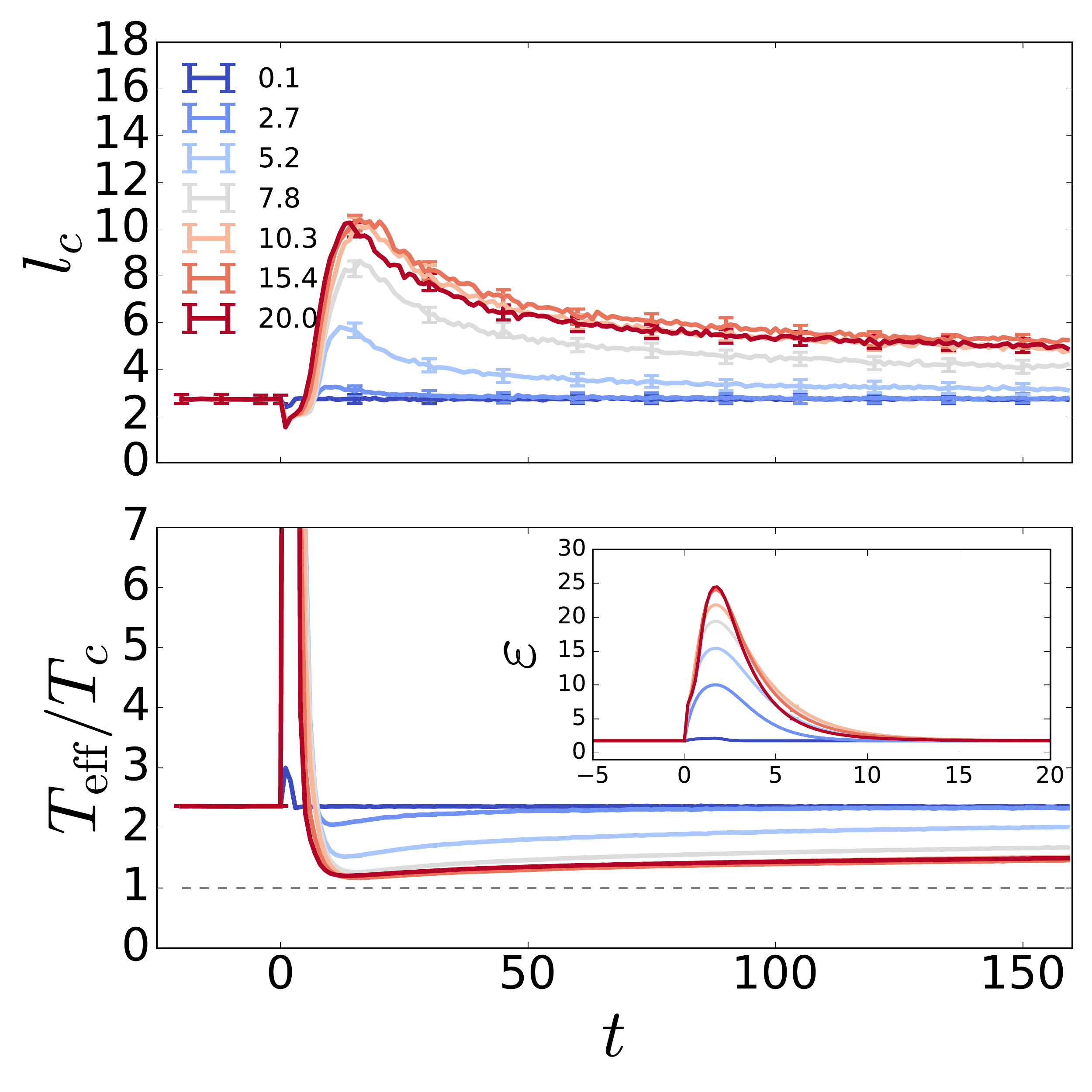}};
    
    \node[align=center,fill=white] at (-3.6, 3.8) {\textbf{a1}};
    \node[align=center,fill=white] at (-3.6, 0.2) {\textbf{a2}};
    
    \end{tikzpicture}
    ~
    \begin{tikzpicture}
    \node[inner sep=0pt] (duck) at (0,0)
    {\includegraphics[width=0.9\columnwidth]{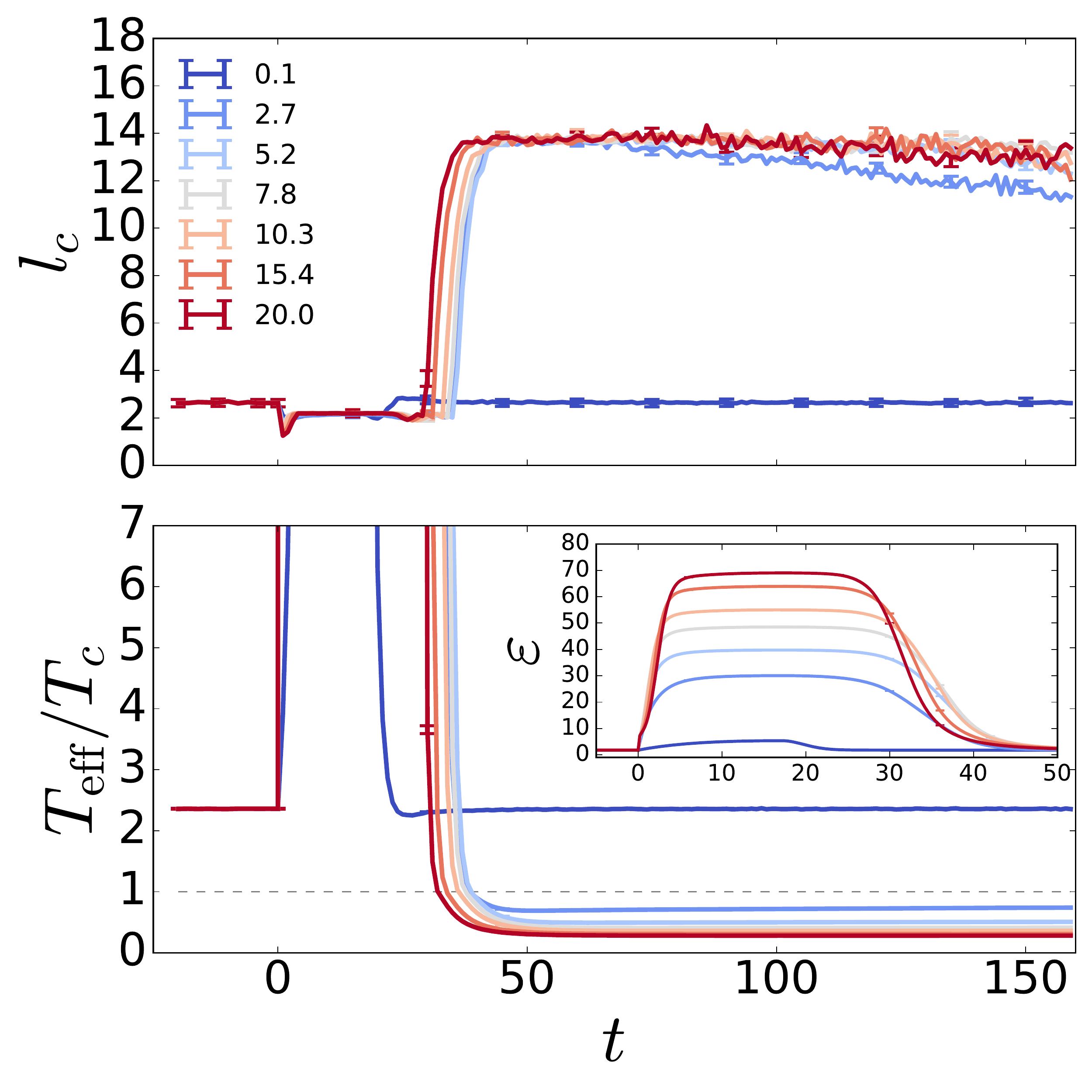}};
    
    \node[align=center,fill=white] at (-3.6, 3.8) {\textbf{b1}};
    \node[align=center,fill=white] at (-3.6, 0.2) {\textbf{b2}};
    
    \end{tikzpicture}

    
        \caption{
\label{fig::evolution_energy_OP}
Correlation length 
$l_c(t)$ of transient $U(1)$ ordering for shorter quenches (a1) and longer quenches (b1). The quench strength parameter $\kappa$ is given in the plot legends.  
 (a2, b2)~The  effective
temperature $T_{\rm eff} (t)$ for cold sites corresponding to plots (a1) and (b1) respectively. The gray dashed line indicates  $T_{\rm eff} = T_c$. In (a2), $T_{\rm eff}$ stays above $T_c$ but comes close to it. In (b2), $T_{\rm eff}$ 
drops significantly below $T_c$.
Insets of (a2) and (b2): Energy density $\varepsilon(t)$ during the quenches.  
}
\end{figure*}
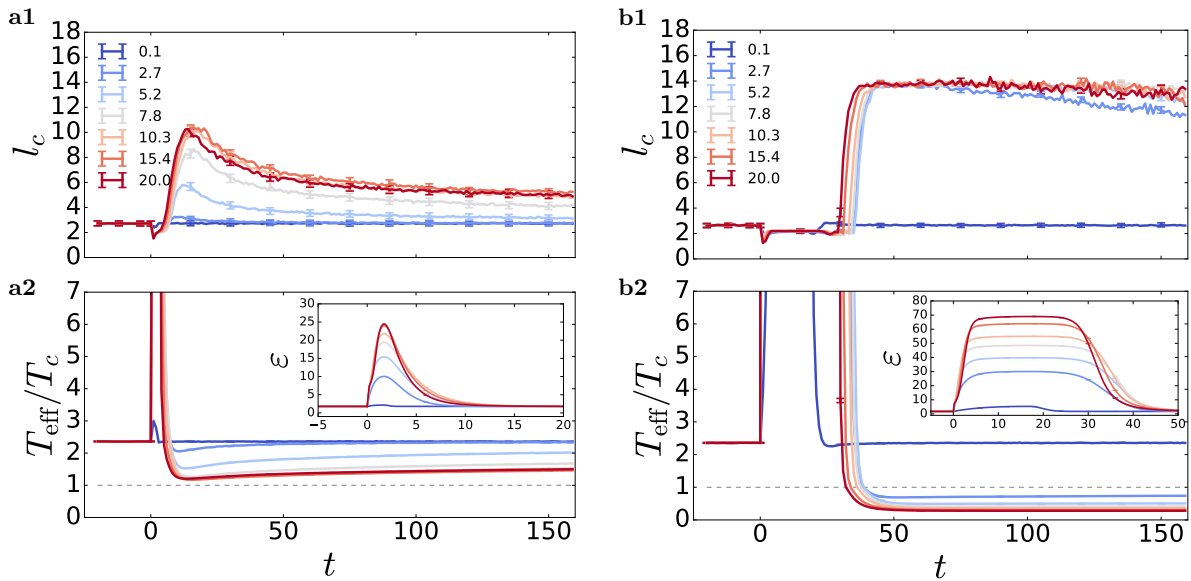

\textit{Transient ordering. ---}
Before the quench, the system is prepared in an equilibrium state in a
non-ordered phase at temperature
$T_0 > T_{\rm c}$.
Then, the system is subjected to a fast energy increase followed by a
cooling step, which brings the energy back to its pre-quench value. This is
achieved by introducing a time-dependent ``gauge-invariant'' term
$-K \psi_{j} D_j$
into the right-hand side of
Eq.~(\ref{eq::effect_dgpe}).
Here
$D_j^{\vphantom{*}} = i
\sum_{m\in \text{NN}(j)} \left( \psi_{m}^* \psi_{j} - \psi_{m} \psi_{j}^*
\right)$,
real function
$K = K(t, \kappa)$
is
designed~\cite{supplemental}
to guarantee that the total energy grows initially but later returns to the
pre-quench value. Parameter $\kappa$ controls the quench
strength~\cite{supplemental}.
Our quench term directly changes only the kinetic
energy of the system, but this change then quickly affects the potential energy through the system's dynamics.

Our main results for various quenches starting from equilibrium at temperature $T_0 = 2.3 \  T_c$ are presented in Fig.~\ref{fig::evolution_energy_OP},
where panel (a1) shows the correlation length $l_c(t)$ for a family of shorter quenches of varying strength, while panel (b1) does the same for a family of longer quenches. The profiles of the quenches are shown in the insets of panels (a2) and (b2) respectively. When $t<0$, $l_c \approx 3$ in units of the  lattice period. Once a quench is launched at
$t=0$, the system energy spikes, the correlation length initially decreases during the heating stage, but then  starts growing during the cooling stage, reaching the maximum around the end of the  cooling and then very slowly relaxing back to the equilibrium value. The maximum value of $l_c(t)$ for the shorter quenches is factor of three larger than the equilibrium one. For longer quenches the increase of $l_c(t)$ is even more significant, but here the maximum of $l_c$ is limited by the size of the simulated lattice ($V=50\times50\times50$), which is means that it would be even larger in the thermodynamic limit.
Overall, the above phenomenology means that the local $U(1)$ ordering exhibits  a large long-lived transient increase. We also note that the longer quenches lead to a significantly longer lifetime of the transient order than the shorter ones.

\textit{The origin of the transient ordering. ---}
Our analysis indicates that the observed transient ordering is caused by the emergence of a small number of ``hot'' lattice sites that have anomalously large norm $|\psi_j|^2$ and thus carry even more anomalous local potential energy $\frac{g}{2}|\psi_j|^4$ --- see the sketch in
Fig.~\ref{fig::sketch_small}.
While making only a small percentage of all lattice sites, the hot sites,
trap a significant fraction of the total energy deposited into the system
during the initial stage of the quench. At the same time,  these sites
become largely decoupled from the rest of the system after the quench and
thus relax very
slowly~\cite{supplemental}.
Since after the quench the total energy of the
system returns to the initial value, the trapping of the energy by the hot
sites implies that the energy density after the quench for the rest of the
system must become smaller than the initial energy density. As a result,
the effective temperature after the quench for the part of the system
excluding hot sites becomes lower than the initial one. This lower
effective temperature gets closer to the ordering temperature $T_c$ and may even become lower than $T_c$  as was the case for our longer quenches. 

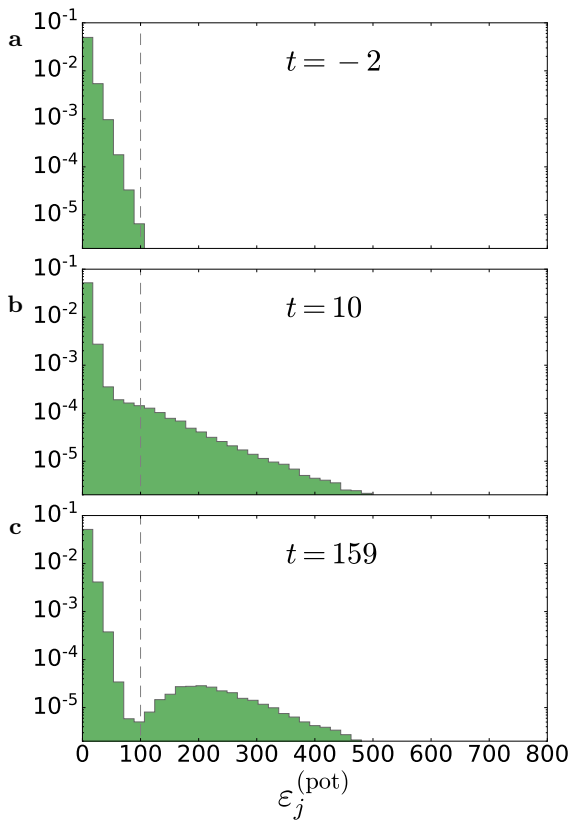
\begin{figure}[t!]
    \centering
    
    \begin{tikzpicture}
    \node[inner sep=0pt] (duck) at (0,0)
    {\includegraphics[width=0.85\columnwidth]{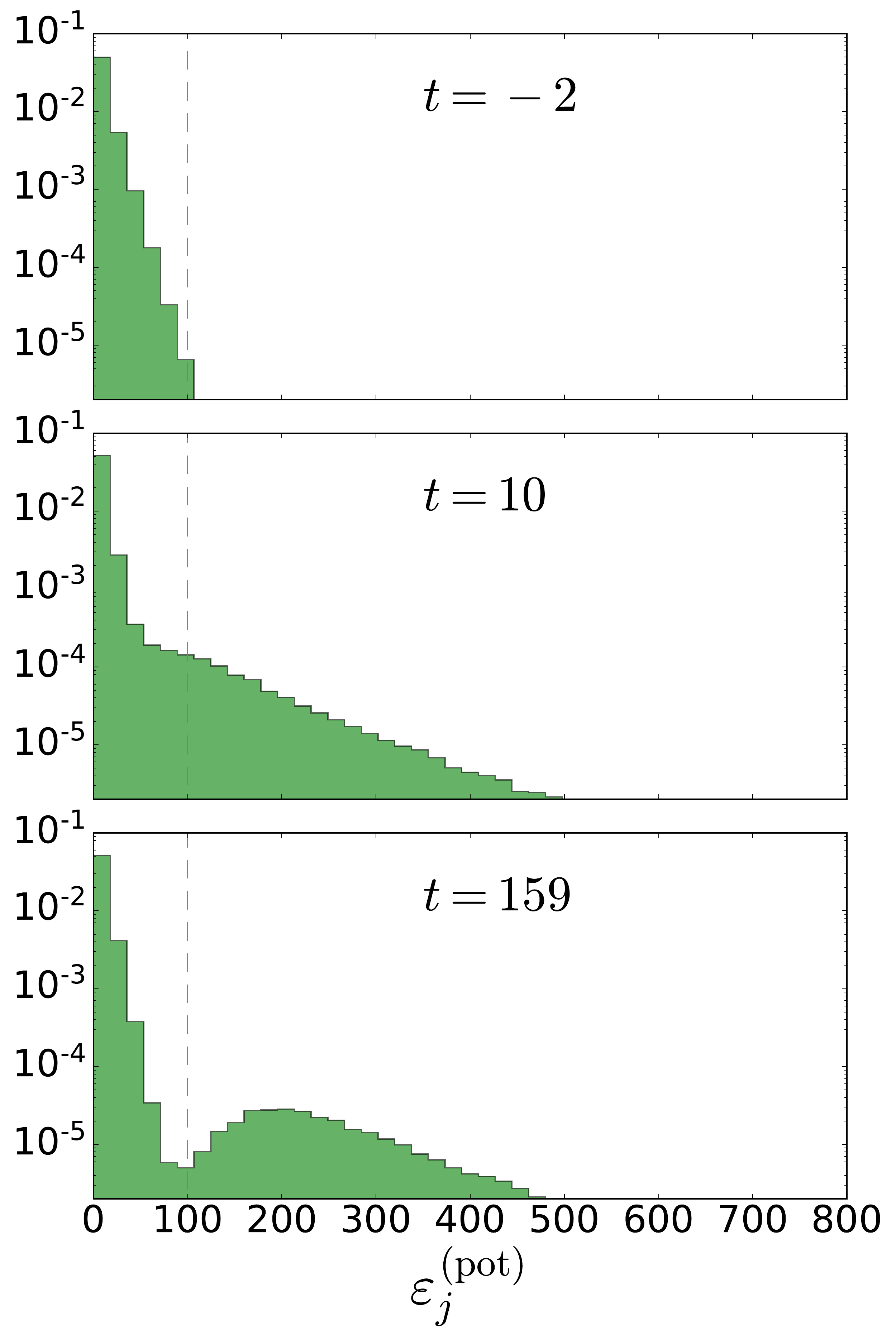}};
    
    \node[align=center,fill=white] at (-3.8, 5.0) {\textbf{a}};
    \node[align=center,fill=white] at (-3.8, 1.5) {\textbf{b}};
    \node[align=center,fill=white] at (-3.8, -1.5) {\textbf{c}};
    
    \end{tikzpicture}

        \caption{
\label{fig::pot_energy_distr}
Histograms of local potential energies
$\varepsilon_j^{\rm (pot)}=\frac{g}{2} |\psi_j|^4$ corresponding to the shorter quench in Fig.~\ref{fig::evolution_energy_OP}(a1,a2) with $\kappa = 20$.
(a)~Equilibrium distribution.
(b)~Distribution immediately after the quench ($t=10$).
(c)~Distribution during the post-quench relaxation ($t=160$).
The horizontal axis represents local potential energy, the vertical axis
shows the fraction of sites with a given value of
$\varepsilon_j^{\rm (pot)}$.
Presence of a larger number of sites with atypically large potential energy
is visible in (b,c). The threshold energy
$\varepsilon_{th}=100$
is marked by vertical dashed lines.
}
\end{figure}

To substantiate the above explanation, we examine the histograms of local potential energies 
$\varepsilon_j^{\rm (pot)}=\frac{g}{2} |\psi_j|^4$ as functions of time. 
These histograms in equilibrium and at two different times after a quench  
are presented in
Fig.~\ref{fig::pot_energy_distr}.
While comparing equilibrium and non-equilibrium histograms, one notices
the salient enhancement of the number of sites with very high values of the
potential energy in far-from-equilibrium states. Energy does not spread
evenly over the whole system, but instead preferably accumulates on a few
sites. 

In order to formally divide the system into ``hot" and
``cold" sites: , we introduce a cutoff for the one-site potential energy
$\varepsilon_{th} \gg E/V$: any site $j$, for which the potential energy
$\varepsilon_j^{\rm (pot)} \equiv \frac{g}{2} |\psi_j|^4  > \varepsilon_{th}$,
is considered hot, otherwise, it is cold. We set
$\varepsilon_{th} = 100$. With such a choice,
the percentage of hot sites $x_{\rm hot}$ for our pre-quench equilibrium state at
$T_0 \approx 2.3 T_{\rm c}$ is nearly zero,  and so is their total potential energy $E_{\rm hot}^{\rm (pot)}$.
The quench acts to increase both $x_{\rm hot}$ and $E_{\rm hot}^{\rm (pot)}$. For the state represented
in
Fig.~\ref{fig::pot_energy_distr}(c),
one has
$x_{\rm hot} \approx 0.5 \%$
and
$E_{\rm hot}^{\rm (pot)}/E \approx 0.7$.
As a result, the energy density of cold sites drops significantly below the initial energy density
$\varepsilon$. Since the fraction of hot sites
$x_{\rm hot}$
is minuscule, the ensemble of the cold sites essentially coincides with the
whole system. 

To demonstrate the overcooling of the cold sites, we
monitor their effective temperature
$T_{\rm eff}$. As ``a thermometer" for the cold subsystem, we use the kinetic energy
density of the cold sites obtained as $\varepsilon^{\rm (kin)}_{\rm cold} (t)
=
\frac{E^{\rm (kin)}(t) - E^{\rm (kin)}_{\rm hot}(t)}
	{\left(1 - x_{\rm hot}\right) V},$ where $E^{\rm (kin)}$ is the total kinetic energy of the lattice and
$E^{\rm (kin)}_{\rm hot}$ is the part involving the hot sites.
The values of $\varepsilon^{\rm (kin)}_{\rm cold}$ are then converted into $T_{\rm eff}$
using the computed equilibrium plot of temperature as a function of
$\varepsilon^{\rm (kin)}$ given in \cite{supplemental}. The resulting
dependence $T_{\rm eff}(t)$ is presented in
Fig.~\ref{fig::evolution_energy_OP}(a2,b2). As expected, one sees there that, for both shorter and longer quenches,
$l_c(t)$ grows as $T_{\rm eff}(t)$ decreases towards $T_c$, which  indicates that the transient
ordering originates from the overcooling of the cold part of the system.

\textit{Discussion and conclusions. ---}
The DGPE on a sufficiently large lattice cluster represents a very simple
model of a microcanonical thermodynamic system exhibiting a phase
transition into a
$U(1)$-ordered 
phase. If a time-dependent perturbation, representing external drive, is
included, one can use the DGPE to study non-equilibrium dynamics near the
continuous transition. The model is characterized by the following three
features: (i)~in equilibrium, the model has a single ordered phase; (ii)~by the
model's very design, post-quench dynamics is inseparable from equilibrium
state properties, as both are governed by the same set of differential
equations; (iii)~all equilibrium and post-quench properties are controlled
entirely by the energy density
$\varepsilon$
and the interaction parameter $g$. Points (i-iii) imply that very little
room for fine-tuning is available for the DGPE. Despite that, however, the
model demonstrates a robust transient ordering, which is a remarkable non-equilibrium
phenomenon. 


In the simulations, the transient ordering manifested itself as a dramatic increase of the phase coherence length $l_c$. This increase  
persists much longer than the duration of the quench; its relaxation is then
remarkably sluggish [see
Fig.~\ref{fig::evolution_energy_OP}(a1,b1)] --- consequence of the long lifetime of the hot sites. The emergence of hot sites is the property of the high-energy regime of DGPE, which is known to exhibit poorly ergodizing dynamics. Reaching high energies of the DGPE lattice necessarily requires the potential energy of the system to become high, but this is only possible when the distribution of norm $|\psi_j|^2$ becomes highly inhomogeneous, thereby facilitating the large energy contribution $\frac{g}{2}|\psi_j|^4$ from the hot sites.  The strong energy quench just takes the system into that poorly ergodizing  regime, and then the hot sites remain ``stuck'' in that regime after the energy density of the system returns to the initial value. 

We remark in this regard that the strength of the quench in the present investigation is significantly larger than in our closely related work~\cite{tarkhov2022dynamics}, where the focus was on investigating vorticity around the phase transition temperature, hence the quenches did not reach the poorly ergodizing regime. Another remark is that our numerical implementation of the quench was based on pumping the kinetic energy associated with the terms $\psi_m^* \psi_j$. This procedure is uniform throughout the system, and then the hot sites arise dynamically. We suppose that the transient ordering would also be achieved by pumping the potential energy, but making the potential energy very high is only possible by distributing the norm $|\psi_j|^2$ nonuniformly, which would introduce an additional arbitrary element into our simulations.

The experimental context of our simulations extends to physical systems exhibiting phase transitions associated with $U(1)$ ordering. These, in particular, include superfluid and superconducting systems, systems exhibiting density-wave orders, and also magnetically-ordered systems with easy-plane anisotropy. One can coarse-grain these systems into parts that are smaller than the expected length of the $U(1)$ phase coherence and yet large enough to justify the classical modelling of each coarse-grained element.  Each variable $\psi_j$ of our modelled lattice would  then be associated with the average $U(1)$ order within one coarse-grained element (see e.g., Ref.~\cite{tarkhov2022dynamics}). 
The quench can be implemented in solid-state systems by a fast heating of the system by a laser pulse, followed by fast cooling of the excited degrees of freedom either by a much larger heat reservoir or just by turning out the laser pulse. 

Several
cases of the transient $U(1)$ ordering 
have already been reported in the literature~\cite{ligh_ind_supercond_cuprat2011exper,non_eq_supercond2015exper,
mitrano2016possible, cantaluppi2018pressure,budden2021evidence}. 
Particularly relevant here are the observations in the alkali-doped fulleride superconductor K$_3$C$_{60}$ \cite{budden2021evidence}. The system there was excited at temperatures significantly above the superconducting $T_c$, and then not only a superconducting-like response was observed as such, but also its lifetime has become dramatically longer once the energy pumped by the laser pulse increased. In other words, a stronger heating resulted in a stronger low-temperature-like response, which is reproduced by our simulations (compare shorter quenches in Fig.~\ref{fig::evolution_energy_OP}(a1,a2) to longer quenches in Fig.~\ref{fig::evolution_energy_OP}(b1,b2)).
 Although modelling of a real superconducting system based on DGPE is a rather oversimplified approach, it is difficult to ignore the remarkable parallels between the above experiment and our simulations. In the simulations, the attainment of the transient ordering required us to pump a lot of energy into the system to reach the regime, where the potential energy can no longer be uniformly distributed over the lattice, which, in turn, led to poorly ergodized dynamics characterized by the long-living hot sites. This suggests a generic mechanism of transient ordering in real systems, namely, (i) an energy quench brings the system into a poorly ergodizing regime, and then (ii) the ordering emerges as a consequence of the dynamical memory  of that regime. We note that the long life of the hot DGPE sites is simply the consequence of their dynamical decoupling from the ``normal'' sites. Such a behavior can be reasonably expected not only for the on-site potential energy of the form prescribed by DGPE but also for a broader class of potential energies. 
 
 Our treatment of transient coherence can be compared with alternative ways of theoretical modeling of light-induced phase coherence. We note in this regard that 
 the long life of the superconducting response in the experiment of Ref.\cite{budden2021evidence} after the end of the laser pulse is a challenge for the  theoretical approaches explaining the laser-induced superconductivity by a periodic driving of the electron-phonon system\cite{raines2015,hoppner2015,denny2015,komnik2016,murakami2017,kennes2017,babadi2017,dai2021}. Whether the long-lived transient coherence can be explained using explicitly dissipative modelling, such as done, e.g., in \cite{budden2021evidence,dolgirev2021periodic}, is yet to be seen.



To conclude, we numerically studied the non-equilibrium evolution of the 3D
DGPE model subjected to an energy quench. Transient $U(1)$ ordering was
consistently observed, provided that the quench was strong enough. We explain this phenomenon in terms of long-living hot
breather-like lattice sites possessing anomalously large potential energies. Such non-equilibrium behavior may
be an intrinsic feature of a broad class of dynamical models. Our
findings may, in particular, shed light on the experimental observations of the transient ordering in superconducting
and charge-density-wave systems.

{\it Acknowledgments. ---} This work was supported by a grant of the Russian
Science Foundation (Project No. 17-12-01587).

{\it Code availability. ---} The code is publicly available in a GitHub repository at \href{https://github.com/TarkhovAndrei/DGPE}{https://github.com/TarkhovAndrei/DGPE}.

\bibliographystyle{apsrev4-1}
\bibliography{refs}

\renewcommand{\theequation}{S.\arabic{equation}}
\setcounter{equation}{0}

\renewcommand{\thefigure}{S.\arabic{figure}}
\setcounter{figure}{0}
\newpage

\section*{Supplemental Material}

\section{Microcanonical properties of DGPE}

For vanishing disorder, ``macroscopic features'' of the DGPE can be
described in terms of a microcanonical ensemble of all phase space states
with fixed $E$ and $N$. The microcanonical temperature is defined as:
\begin{equation}
\label{Intro:eq:microcan_T_def}
\beta \equiv \frac{1}{T} \equiv \frac{\partial S}{\partial E},
\end{equation}
where $S$ is the microcanonical entropy proportional to the logarithm of
the volume of the energy shell
$w(E)$,
\begin{eqnarray}
	S \equiv \log \left[
		\frac{w(E)}{(2\pi\hbar)^{\frac{2V-1}{2}}}
	\right],
\end{eqnarray} 
the Boltzmann's constant is set to be equal to unity, $k_B\equiv 1$. In microcanonical thermodynamics, the dimensionality of energy shell is $2V-1$, however, for a thermodynamically large system the difference between $2V$ and $2V-1$ can be neglected. The original definition of microcanonical temperature  from conservative dynamics for an ergodic system reads~\cite{rugh1997dynamical, rugh1998geometric,den1998calculation, den2000thermodynamic}:

\begin{equation}
\frac{1}{T(E)} \equiv \lim_{t \to \infty} \frac{1}{t}\int_0^t d\tau \Phi(\mathbf{R}(\tau)),\label{Intro:eq:dynamical_T_def_Rugh}
\end{equation} where $\mathbf{R}(\tau)$ is a phase space trajectory and the observable
\begin{equation}
\Phi \equiv \nabla \left( \frac{\nabla \mathcal{H}}{\| \nabla \mathcal{H} \|^2} \right),\label{Intro:eq:dynamical_Phi_functional_def_Rugh}
\end{equation} is representative of the geometric curvature of the Hamiltonian on the energy shell. For the practical calculation of microcanonical temperature, we adapt a numerical recipe, developed for classical spin lattices with one integral of motion~\cite{de2015chaotic} and consistent with the analytical approach~\cite{rugh1997dynamical, rugh1998geometric}, to the calculation of the microcanonical temperature of the DGPE lattices with two integrals of motion. The original idea is to replace the functional in Eq.~(\ref{Intro:eq:dynamical_Phi_functional_def_Rugh}) with its approximate numerical value by sampling the vicinity of the point in phase space $\mathbf{R}(\tau)$, and generating an ensemble of energy realizations, corresponding to each sample point. Then, by calculating the variance of energy fluctuations $\langle\Delta E^2\rangle$ and the mean fluctuation $\langle\Delta E\rangle$, one can extract the local approximation to $\Phi(\mathbf{R})$ as

\begin{equation}
{\Phi}(\mathbf{R}) \equiv \frac{2\langle\Delta E\rangle}{\langle\Delta E^2\rangle}.\label{Intro:eq:dynamical_T_def_Fine_Wijn}
\end{equation} 
We note that $\langle\Delta E\rangle \neq 0$ due to the fact that, in
general, there are exponentially more states above the energy shell than
below it, which in turn reflects the fact that the entropy in
Eq.~(\ref{Intro:eq:microcan_T_def})
is proportional to the logarithm of the energy shell volume, and the
temperature is positive. The temperature is then calculated as the time
average along the phase space trajectory of
Eq.~(\ref{Intro:eq:dynamical_T_def_Fine_Wijn}), $1/T = \bar{\Phi}$. 

We use
Eq.~(\ref{Intro:eq:dynamical_T_def_Fine_Wijn})
with an additional constraint due to the fixed second integral of motion,
the number of particles, hence, when sampling the vicinity of a phase space
point, we sample only the vicinity on the shell of a fixed number of
particles,
$\langle \ldots \rangle_N$, which gives
\begin{equation}
T = \frac{\langle\Delta E^2\rangle_N}{2\langle\Delta E\rangle_N}.
\label{Intro:eq:T_E_fixed_N}
\end{equation}
This equation acts as a practical definition of microcanonical temperature. It
satisfies all properties expected of temperature, and can be implemented
numerically for a DGPE system. We will use 
Eq.~(\ref{Intro:eq:T_E_fixed_N})
to characterize DGPE equilibrium.

\section{Energy quench\label{SM:section:quench}}

The quench is performed on a DGPE system prepared in equilibrium. During
the quench, the system's energy experiences quick rise and fall within a
limited time interval. To allow for time variation of $E$, the DGPE must be
modified to include a time-dependent term:
\begin{eqnarray} 
\label{eq::dgpe_quench}
i\frac{d\psi_{j}}{dt}
=
-\sum_{m\in \text{NN}(j)}
	\psi_{m} + g \left|\psi_{j}\right|^{2}\psi_{j} 
	-  K \psi_{j} D_j,
\\
\text{where}\quad
D_j^{\vphantom{*}} = D_j^* = i \sum_{m\in \text{NN}(j)}
\left( \psi_{m}^* \psi_{j} - \psi_{m} \psi_{j}^* \right),
\end{eqnarray} 
and
$K = K(t, \kappa)$
is a real-valued function. The last term in
Eq.~(\ref{eq::dgpe_quench})
is introduced to control the variation of $E$ during the quench. Note that
Eq.~(\ref{eq::dgpe_quench}),
being gauge invariant, conserves $N$. The structure of the pump term is
complicated and does not lend itself to immediate intuitive interpretation.
To appreciate it, we can calculate
$\dot{E}^{\rm (kin)}$
and
$\dot{E}^{\rm (pot)}$
for dynamical
equations~(\ref{eq::dgpe_quench}),
and convince ourselves that 
$\dot{E}^{\rm (kin)}$
acquires additional contribution
$-K \sum_j D_j^2$,
while the expression for
$\dot{E}^{\rm (pot)}$
is unaffected by the pump term. Thus, a quench described by
Eq.~(\ref{eq::dgpe_quench})
impacts primarily the kinetic energy. Of course, since the system allows
exchange between the two types of energy, altering
$E^{\rm (kin)}$
ultimately changes
$E^{\rm (pot)}$
as well. In the main text, we argue that our conclusions about the transient ordering
holds true if we pump
$E^{\rm (pot)}$
instead of
$E^{\rm (kin)}$.

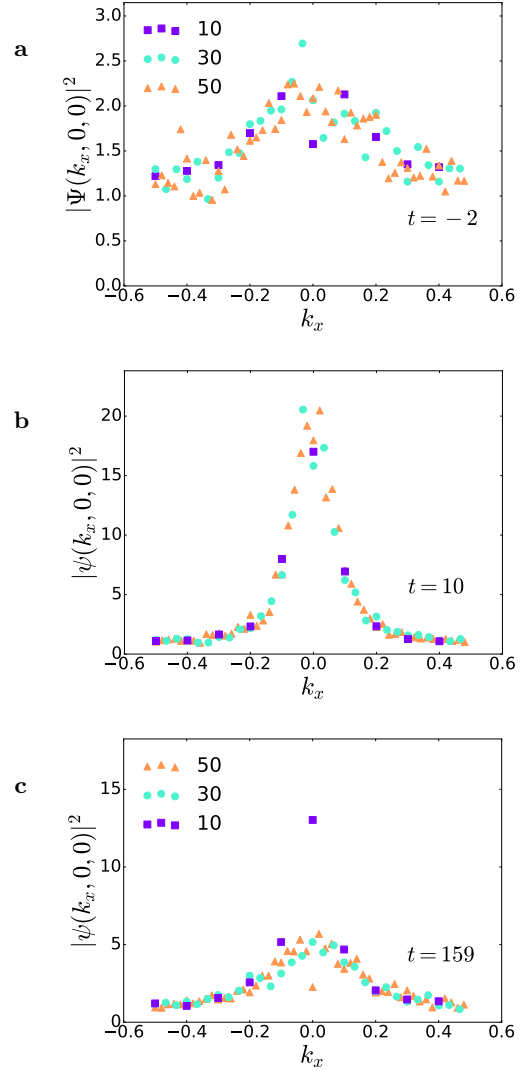
\begin{figure}[t!]
    \centering
    \begin{tikzpicture}
    \node[inner sep=0pt] (duck) at (0,0)
    {\includegraphics[width=0.75\columnwidth]{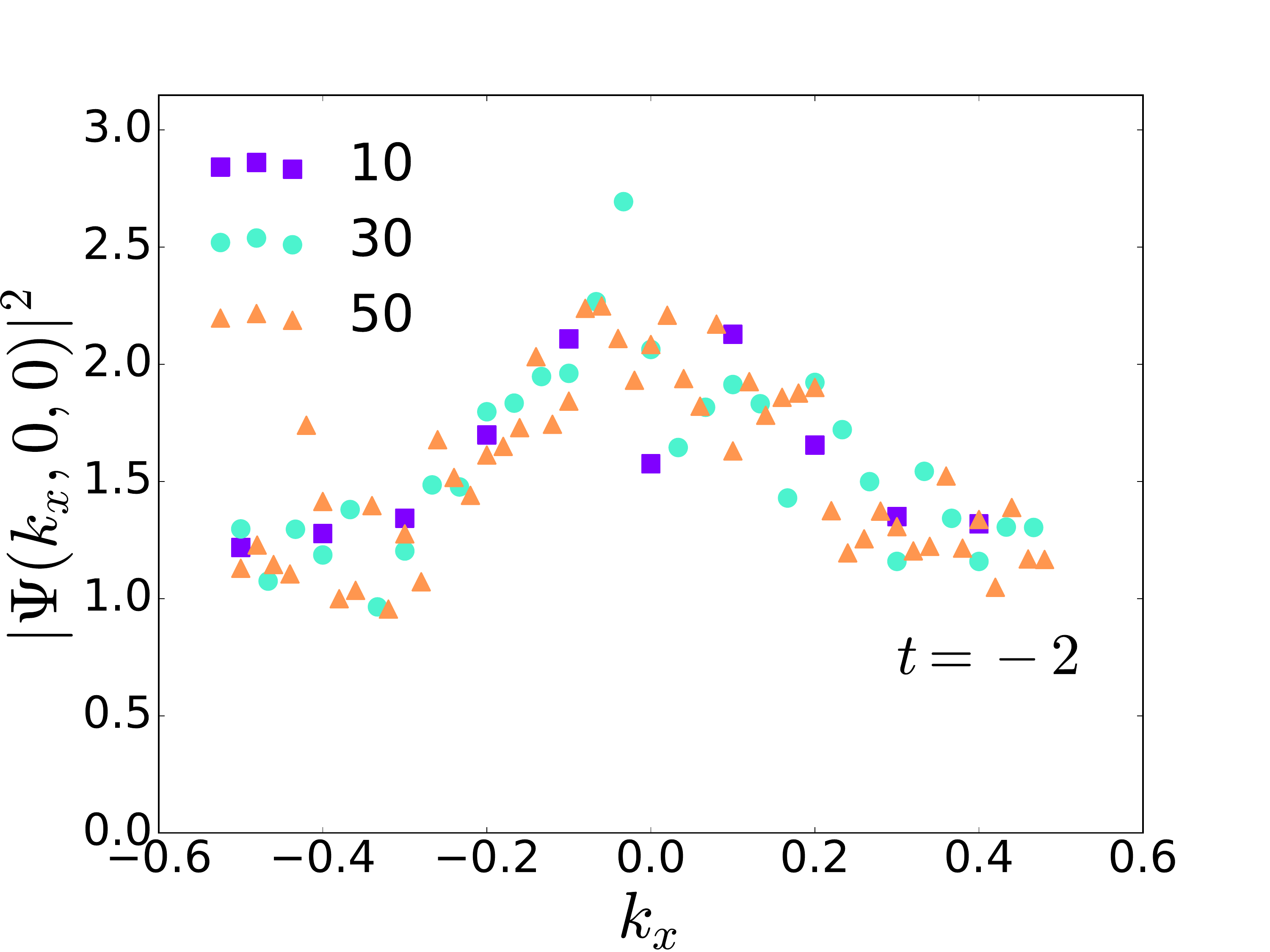}};
    \node[align=center,fill=white] at (-3.8, 1.3) {\textbf{a}};
    \end{tikzpicture}
    
	\begin{tikzpicture}
    \node[inner sep=0pt] (duck) at (0,0)
    {\includegraphics[width=0.75\columnwidth]{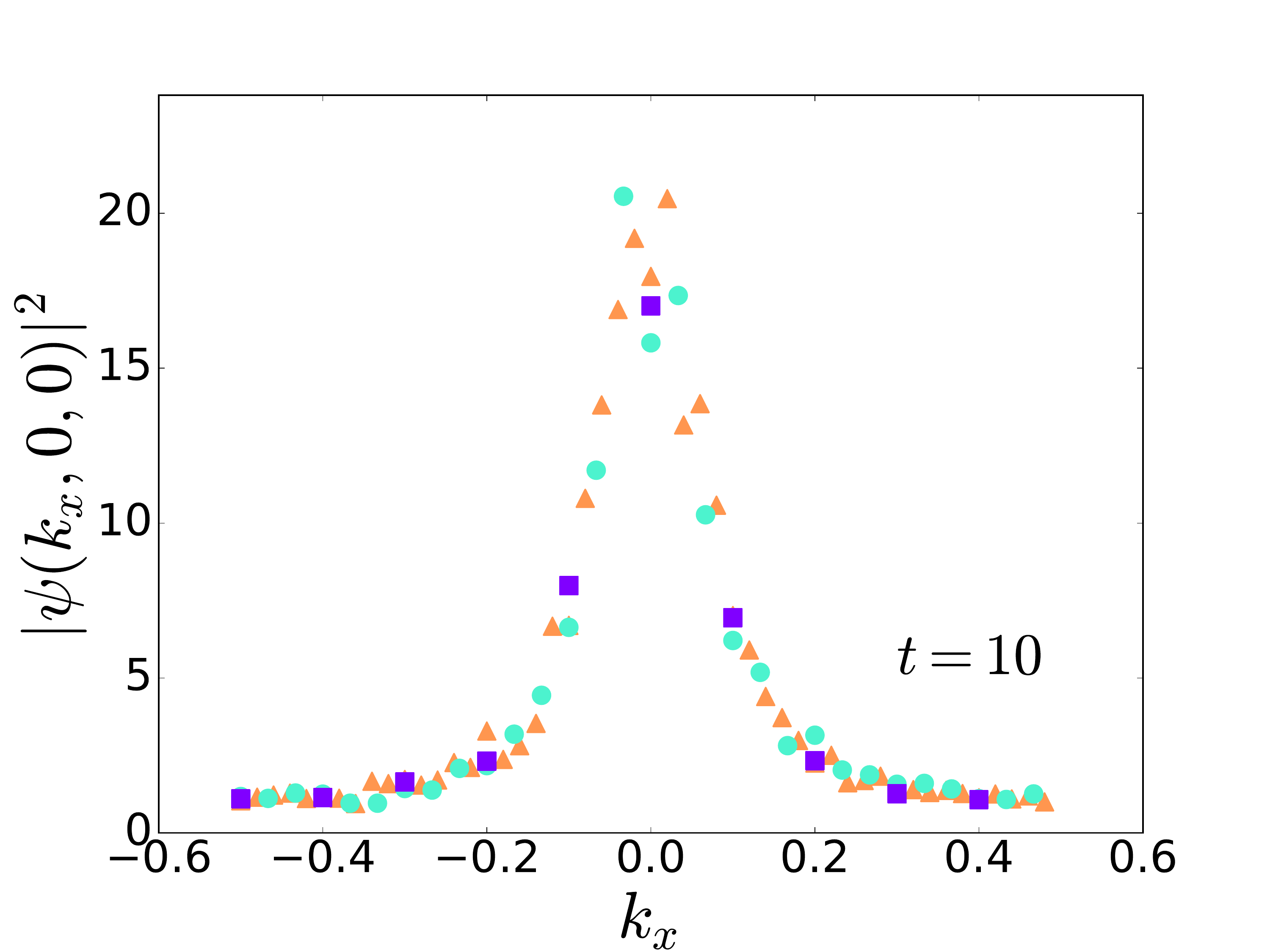}};
    \node[align=center,fill=white] at (-3.8, 1.3) {\textbf{b}};
    \end{tikzpicture}
    
    \begin{tikzpicture}
    \node[inner sep=0pt] (duck) at (0,0)
    {\includegraphics[width=0.75\columnwidth]{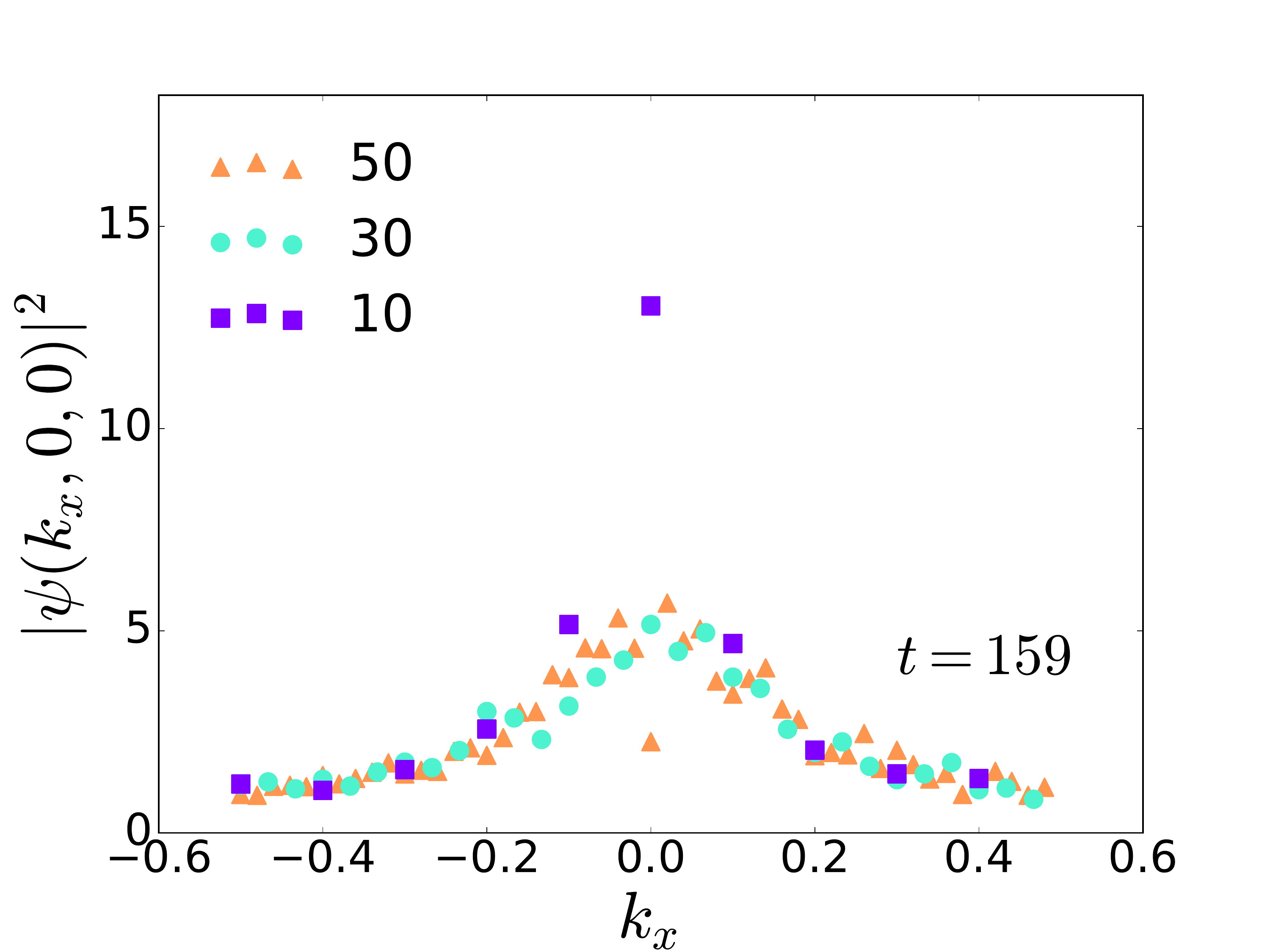}};
    \node[align=center,fill=white] at (-3.8, 1.3) {\textbf{c}};
    \end{tikzpicture}
    
    \caption{
\label{fig::SM_fourier_slices}
The slices of Fourier spectral intensities
$|\psi ({\bf k})|^2$
(a) in equilibrium at
$t = -2$,
and after the short quench: (b) at
$t=10$
and (c) 
$t=159$.
The intensity is plotted against
$k_x$,
while
$k_{y,z}$
are fixed equal to zero. The simulation is performed for 
$g=10$,
$V=10\times10\times10$, 
$V=30\times30\times30$,
and 
$V=50\times50\times50$,
pre-quench temperature
$T_0/T_{\rm c} \approx 2.3$,
quench intensity
$\kappa = 20$.
This figure shows that the correlation length is
non-monotonic: the width is initially large, but then gets smaller, and
finally it is large again.
}
\end{figure}

Since
$\dot{E} = \dot{E}^{\rm (kin)} + \dot{E}^{\rm (pot)} = - K \sum_j D_j^2$,
we conclude that the sign of
$K(t, \kappa)$
controls the increase/decrease of the total energy. Thus, $K$ must be
negative when the system heats, and positive during the cooling stage. We
additionally demand that the energies of the system before and after the
quench are identical. Function
$K(t, \kappa)$
compatible with these requirements is
\begin{eqnarray} 
\label{quench:eq:kappa_profile}
K(t, \kappa)
&=& - \theta (t) \left[
	\kappa \frac{\gamma_2 e^{-\gamma_2 t} - \gamma_1 e^{-\gamma_1 t}}
		{\gamma_2 - \gamma_1}
	\right.
\\
\nonumber 
	&-&	
	\left.
	\theta (t-t^*) \gamma \left(1 - e^{-\gamma_2 (t-t^*)}\right)
	\frac{E(t)-E_0}{E^*-E_0}
\right],
\end{eqnarray} 
where
$\theta (t)$
is the Heaviside step-function, $\kappa$ is the parameter controlling the
quench strength (in our simulations, $\kappa$ spans the range from 0.1 to
20). We used two sets of $\gamma$'s to simulate a short quench ($\gamma_1=1$,
$\gamma_2=0.3$), and a longer quench ($\gamma_1 = 0.03$, $\gamma_2=0.1$); for both quenches $\gamma=0.01$.
The second term in
Eq.~(\ref{quench:eq:kappa_profile})
is added to guarantee that the energy after the quench is the same as
before the quench. Here,
\begin{eqnarray} 
t^*=\frac{1}{\gamma_1-\gamma_2}
	\log \left(\frac{\gamma_1}{\gamma_2} \right)
\end{eqnarray} 
is the time moment at which
$K(t, \kappa)$
crosses zero and changes sign,
$E_0$
is the pre-quench energy of the system, while
$E(t)$
is the energy at time $t$, and
$E^*=E(t^*)$.
For further details one can refer to the source code published in a GitHub
repository~\footnote{The code is published in
a GitHub repository at
\href{https://github.com/TarkhovAndrei/DGPE}
{https://github.com/TarkhovAndrei/DGPE}}.

Before the quench, the system is prepared in an equilibrium low-energy
state. In the first step of the quench the system is subjected to fast
energy increase. The duration of such a pump stage is controlled by 
$\gamma_2$.
In the cooling step (its duration is regulated by 
$\gamma_1$)
the energy is brought back to its pre-quench low value. Energy density
evolution
$\varepsilon = E (t) /V = \varepsilon (t)$
is plotted in the main text.

%
\begin{figure}[t!]
    \centering
    \begin{tikzpicture}
    \node[inner sep=0pt] (duck) at (0,0)
    {\includegraphics[width=0.95\columnwidth]{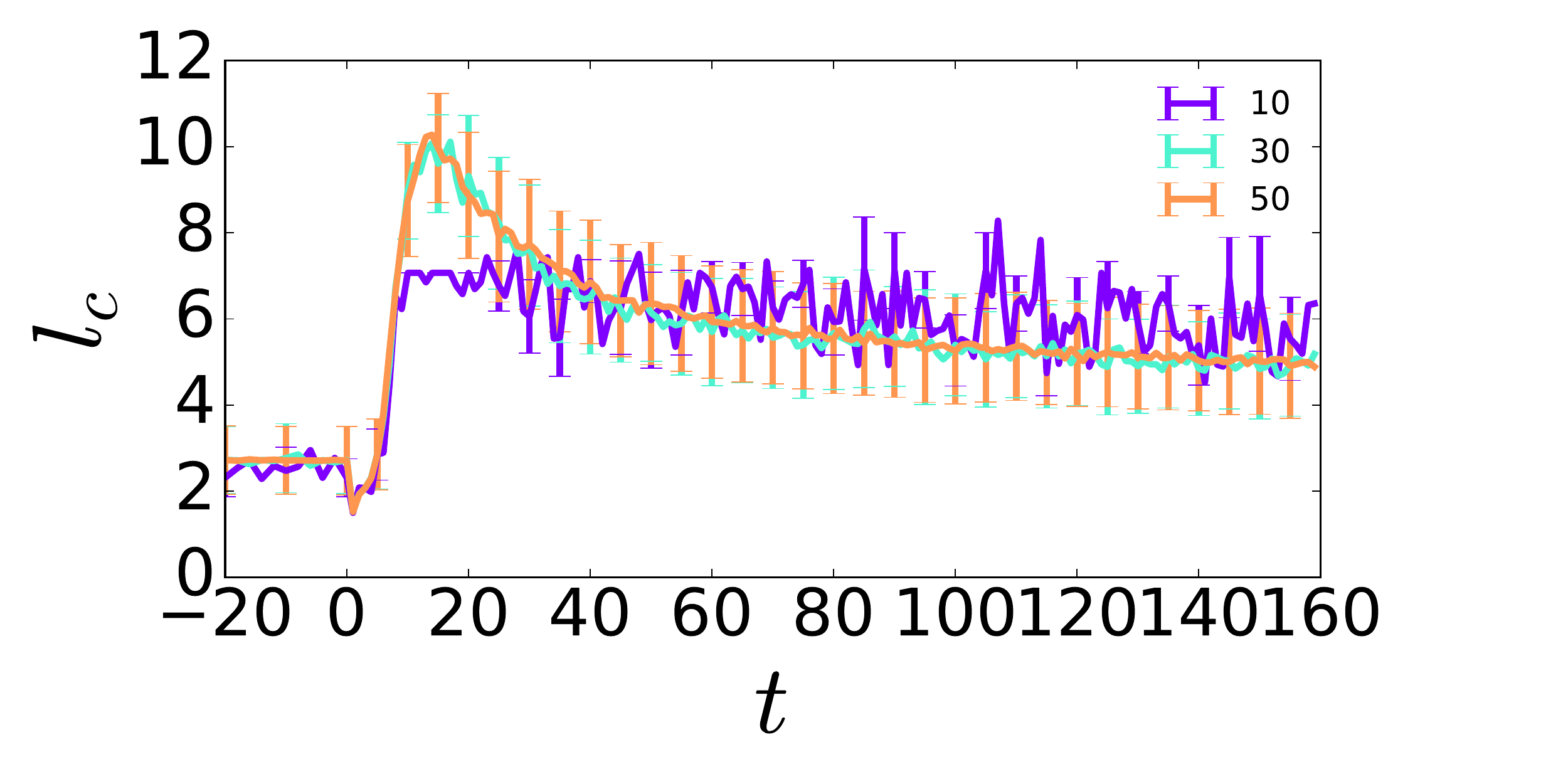}};
    \end{tikzpicture}
    \caption{
\label{fig::lc_finite_size_effects}
The role of finite size effects. The data presents
$l_c(t)$
for systems with $V=10\times 10\times 10$,
$30\times 30\times 30$,
and
$50\times 50\times 50$ subject to the short quench.
For all simulations
$g=10$, $\kappa = 20$.
Finite size effects visibly affect the evolution of
$l_c$ 
in the smallest system, but they are virtually absent for 
$V \geq 30\times 30\times 30$,
where the correlation length remains significantly shorter than the linear
size of the system.
}
\end{figure}

\section{Correlation length and finite-size effects}

\begin{figure}[t!]
    \centering
    \begin{tikzpicture}
    \node[inner sep=0pt] (duck) at (0,0)
    {\includegraphics[width=0.85\columnwidth]{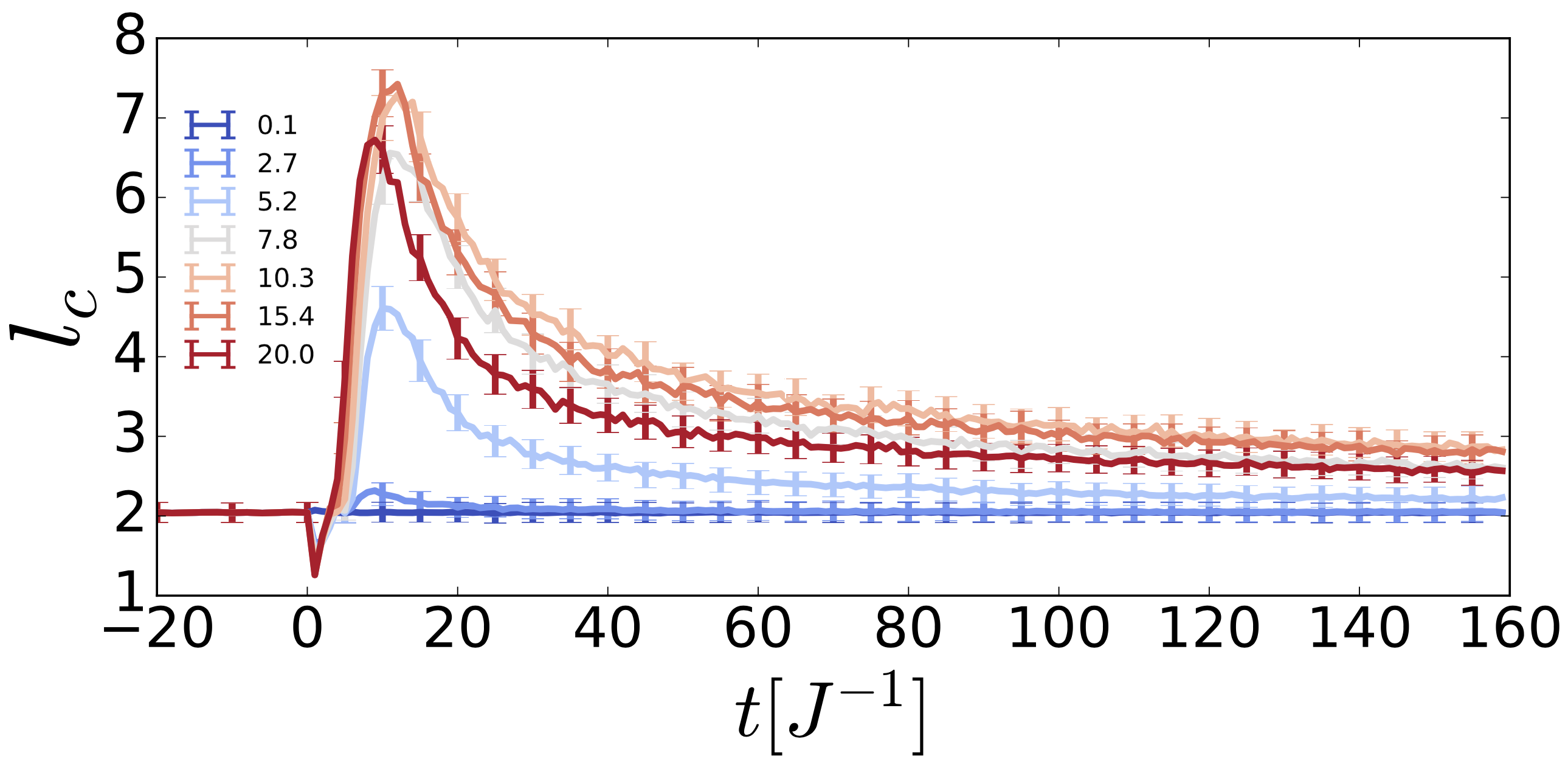}};
    \node[align=center,fill=white] at (-3.8, 1.3) {\textbf{a}};
    \end{tikzpicture}
    
    \begin{tikzpicture}
    \node[inner sep=0pt] (duck) at (0,0)
    {\includegraphics[width=0.85\columnwidth]{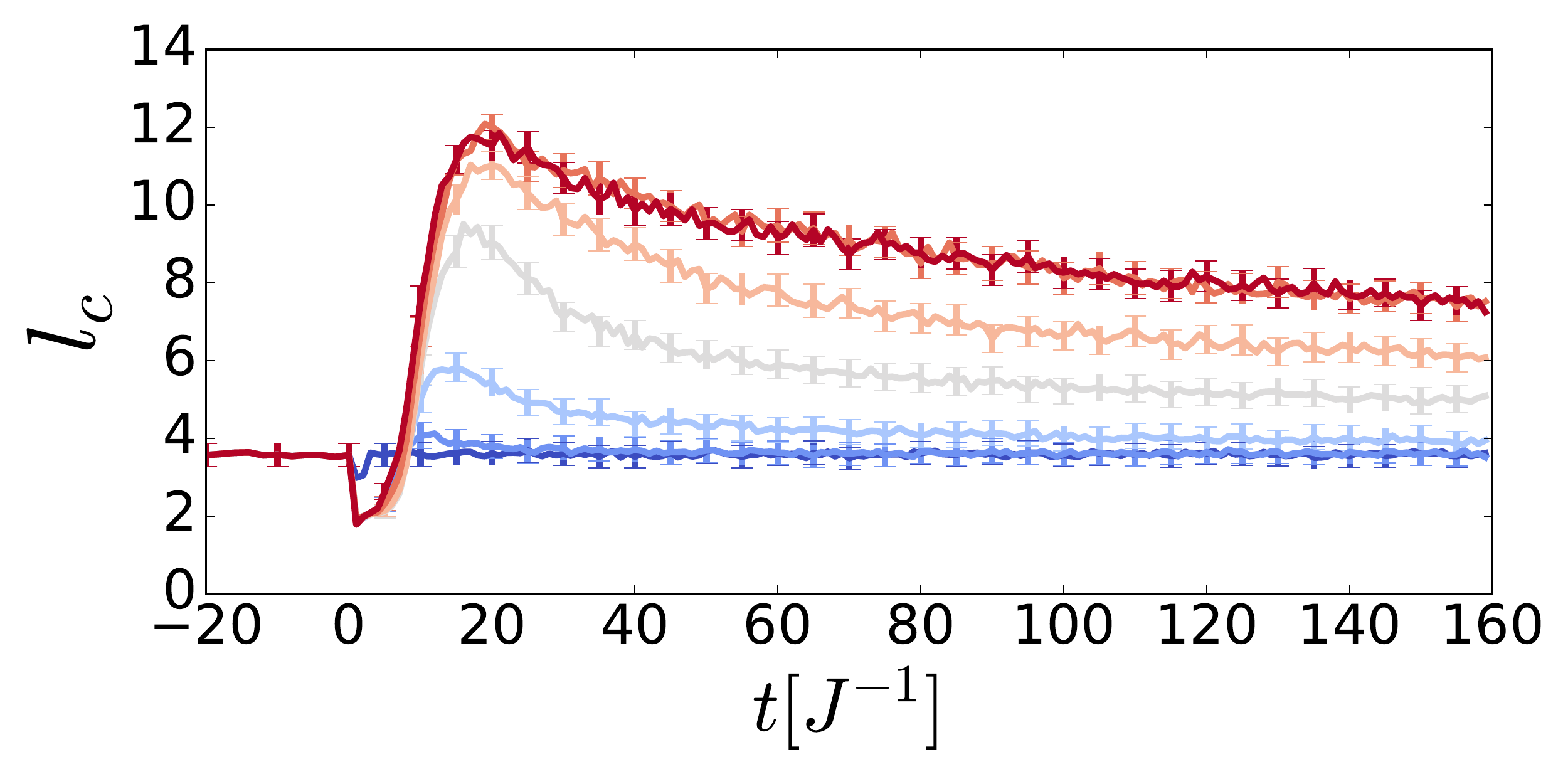}};
    \node[align=center,fill=white] at (-3.8, 1.3) {\textbf{b}};
    \end{tikzpicture}
    
    \caption{
\label{fig::g5_and_g15}
Transient ordering for different interaction strengths $g$.
The short quench parameters are the same as in the main text, but for
(a)~$g=5$,
(b)~$g=15$.
The data for
$g=10$
are presented in
Fig.~\ref{fig::evolution_energy_OP}\,(a1).
For all simulations
$V=50\times 50\times 50$.
The value of $g$ affects the dynamics quantitatively (but not
qualitatively): when $g$ grows $l_c$ becomes larger, and demonstrates
slower relaxation.
}
\end{figure}

In our study, we quantify the transient ordering using the time-dependent
correlation length
$l_c = l_c (t)$,
defined via the half-width at the half-maximum (HWHM) for the spectral
intensity 
$|\psi ({\bf k})|^2$
of $\psi_j$.
The practical calculation of
$l_c$
starts from evaluation of the Fourier transform of 
$\psi_j$
according to the usual prescription
\begin{eqnarray} 
\psi ({\bf k})
=
\frac{1}{V} \sum_{j} 
	e^{i {\bf k} {\bf R}_j} \psi_j.
\end{eqnarray} 
In our simulations we observe that the spectral intensity
$|\psi ({\bf k})|^2$
always has a bell-like shape (with possible exception at
${\bf k} = 0$).
The slices of
$|\psi ({\bf k})|^2$
along 
$k_x$
direction 
($k_{y,z} = 0$)
are shown in
Fig.~\ref{fig::SM_fourier_slices}
for systems of different sizes, both in equilibrium and non-equilibrium
regimes. 
The correlation length defined as
$l_c (t) \equiv 1/\Delta k (t)$,
where
$\Delta k (t)$
is the HWHM for
$|\psi ({\bf k})|^2$
at time $t$, can be extracted with adequate accuracy. For example, a
visual inspection of the three panels of
Fig.~\ref{fig::SM_fourier_slices}
is sufficient to discern the non-monotonic evolution of 
$l_c$
during the quench, consistent with the plots in
Fig.~\ref{fig::evolution_energy_OP}\,(a1).

Besides being a convenient characteristics of the short-range ordering, the
correlation length allows us to assess the severity of finite-size effects
in our simulations. With this goal in mind, let us examine
Fig.~\ref{fig::lc_finite_size_effects}.
There we plot
$l_c (t)$
for three systems of different sizes
($V=10\times10\times10$,  $V=30\times30\times30$, and
$V=50\times50\times50$)
subjected to quench. We see that for
$V=30\times30\times30$
and
$V=50\times50\times50$
the data is essentially insensitive to $V$, and at all times
$l_c$
remains significantly shorter than the linear dimension of the system. This
indicates that, for our purposes, the systems with
$V \geq 30\times30\times30$
are sufficiently large, and the finite-size distortions can be ignored.

Let us finally comment on the relation between 
$l_c$
and the interaction strength $g$. The hot sites of equal norm would accumulate more energy for larger nonlinearity $g$, thus leading to a stronger cooling of the rest of the system after the quench.
Therefore, one may hypothesize that larger $g$ makes the transient ordering stronger.
This intuition is indeed supported by our numerical data, see
Fig.~\ref{fig::g5_and_g15}
and
Fig.~\ref{fig::evolution_energy_OP}(a1).
These two figures present the quench data for three different values of
$g = 5, 10, 15$.
One can see that, when the interaction parameter grows, the correlation
length becomes larger, and remains large for longer time periods.

\section{``Hot" sites: additional characterization}

\begin{figure}[t!]
    \centering
    
    \begin{tikzpicture}
    \node[inner sep=0pt] (duck) at (0,0)
    {\includegraphics[width=0.95\columnwidth]{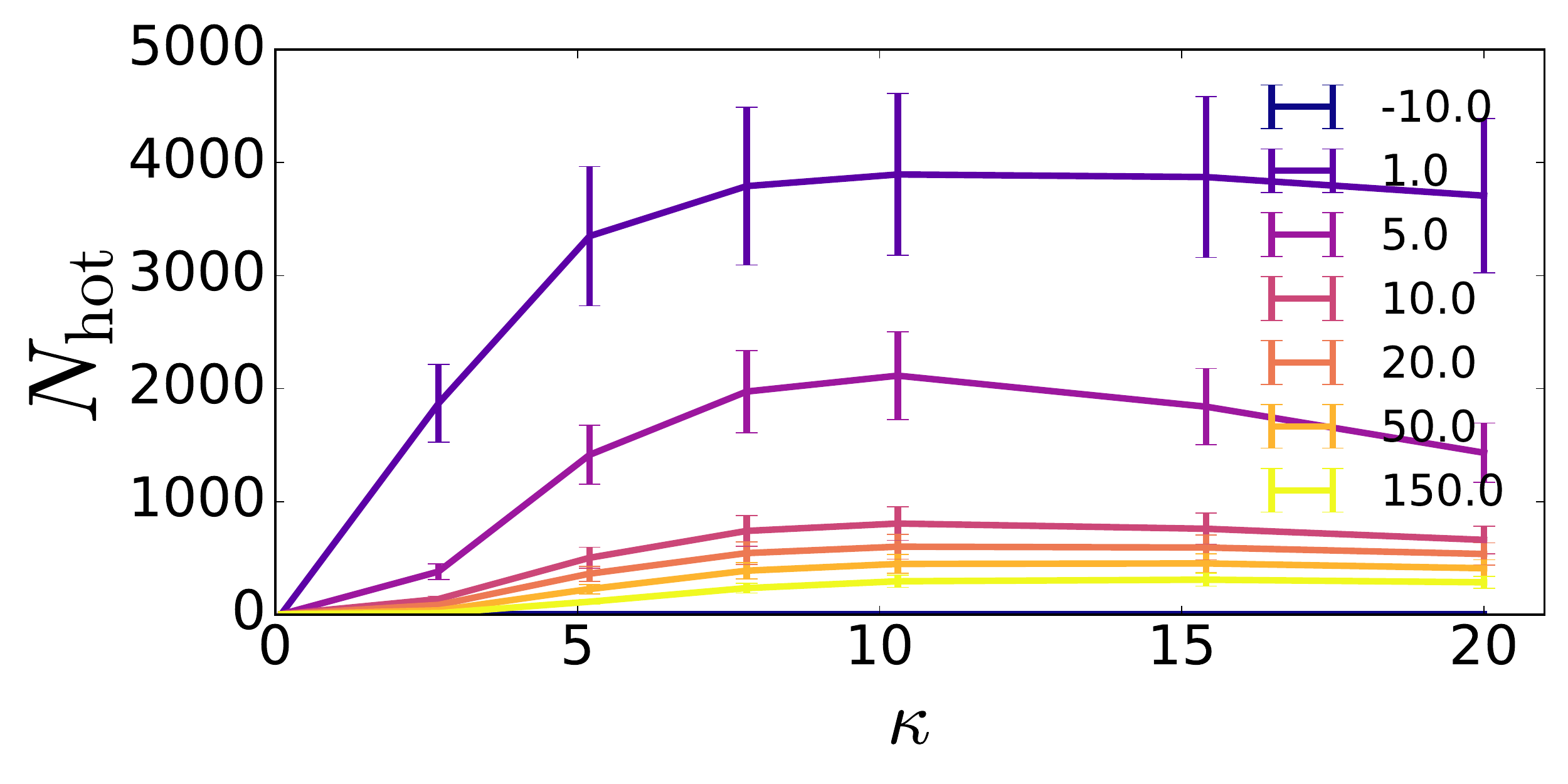}};
    \end{tikzpicture}
    
        \caption{
\label{fig::scaling_of_Nhot_vs_gamma}
Number of hot sites $N_{\rm{hot}}$ as a function of time and quench strength $\kappa$, for $V = 50 \times 50 \times 50$. Before the short quench, in equilibrium, $N_{\rm hot} \sim 0$. The number of hot sites decreases with time, and is maximal for $\kappa \sim 10$.
}
\end{figure}
\begin{figure}[t!]
    \centering
    \begin{tikzpicture}
    \node[inner sep=0pt] (duck) at (0,0)
    {\includegraphics[width=0.85\columnwidth]{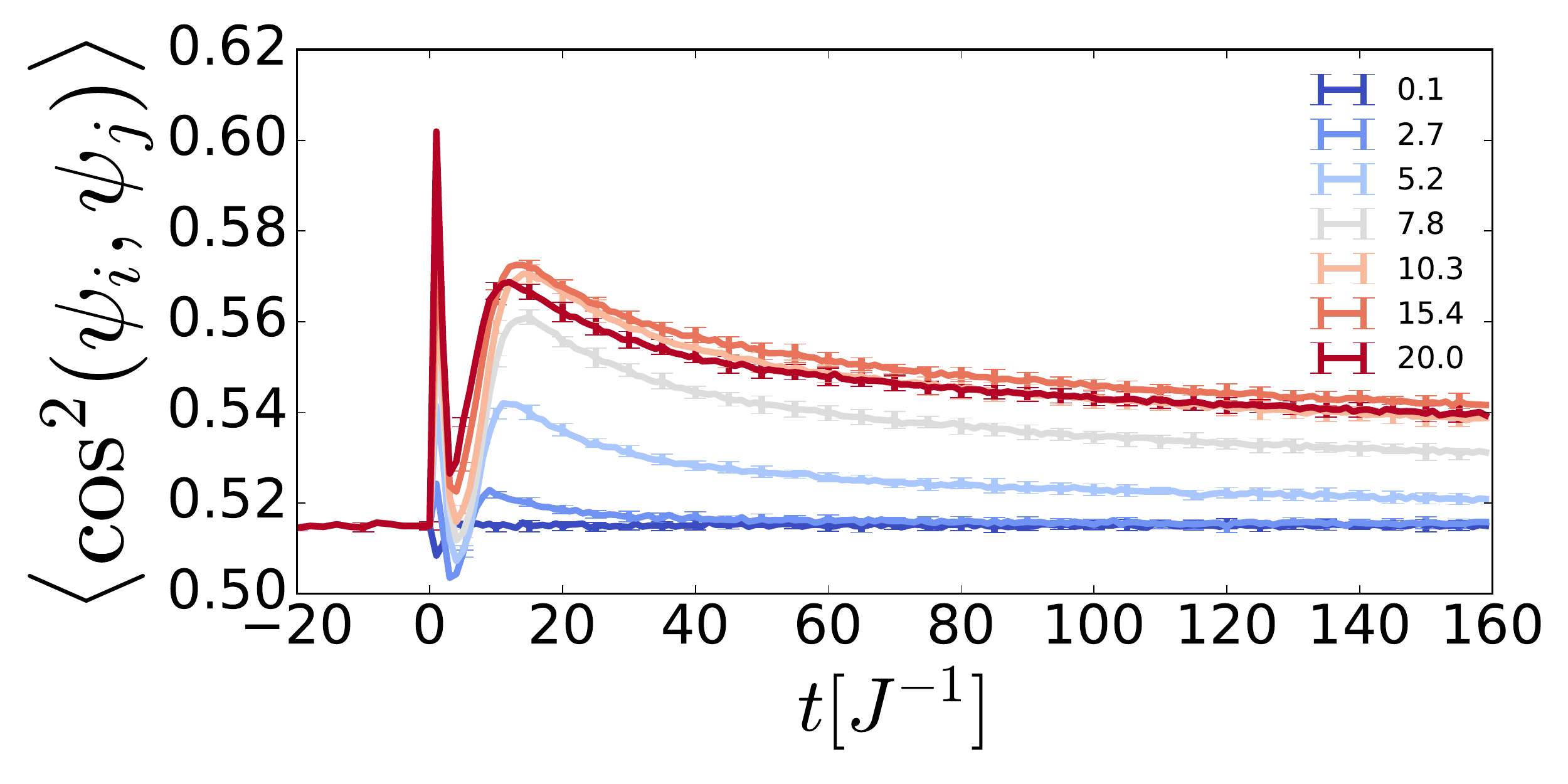}};
    \node[align=center,fill=white] at (-3.8, 1.3) {\textbf{a}};
    \end{tikzpicture}
    
    \begin{tikzpicture}
    \node[inner sep=0pt] (duck) at (0,0)
    {\includegraphics[width=0.85\columnwidth]{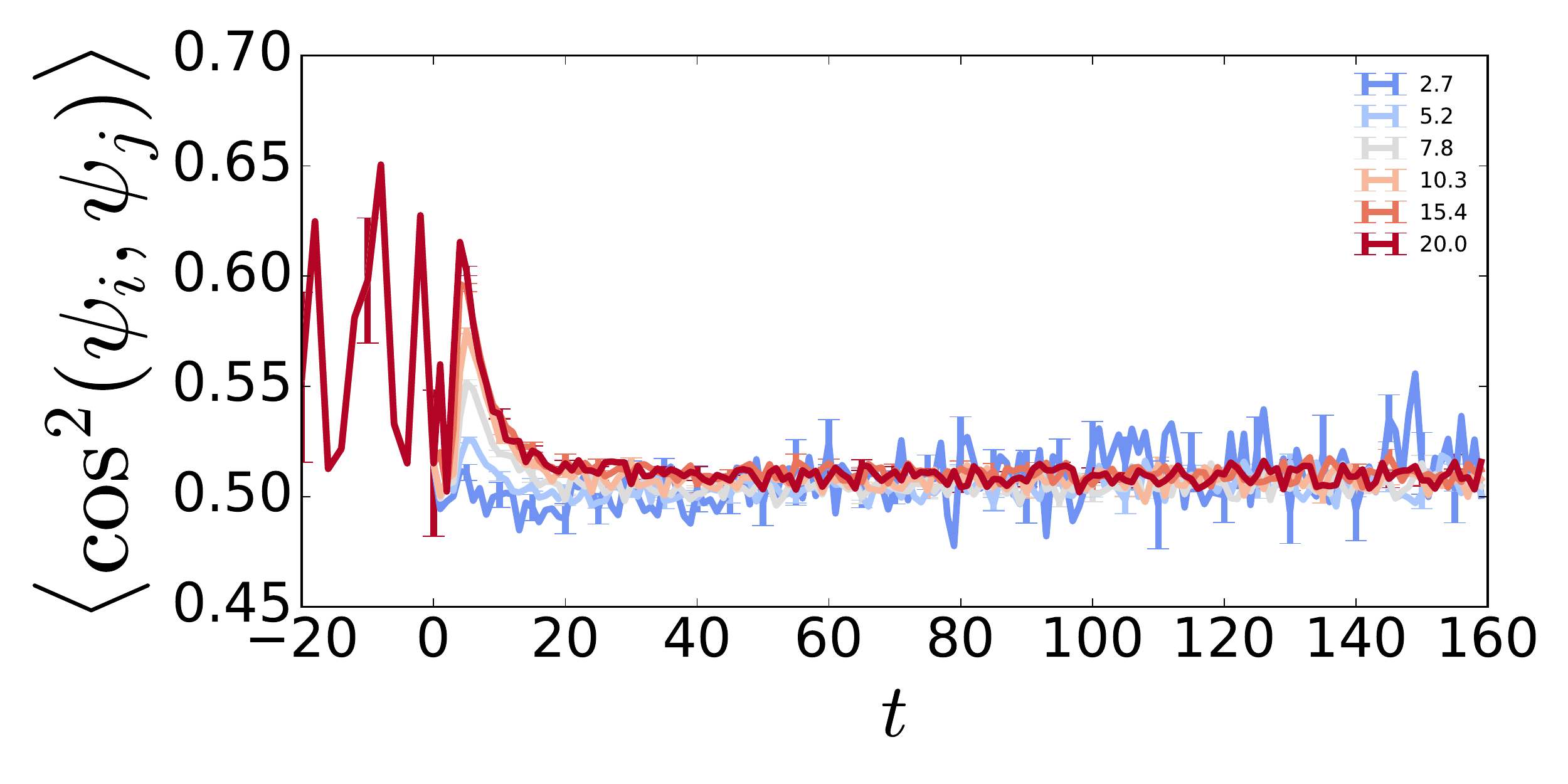}};
    \node[align=center,fill=white] at (-3.8, 1.3) {\textbf{b}};
    \end{tikzpicture}
    
    \caption{
\label{fig::cos_correlation}
The mean of
$\cos^2 (\psi_i, \psi_j)$
for a subset of sites and their neighbors: (a) the cold sites demonstrate a
finite correlation with the neighbors for a long time, whereas (b) the hot
sites quickly decouple from the environment. The simulations were done for
the short quench, $g=10$, $V=50\times 50\times 50$.
}
\end{figure}

As explained in the main text, the transient ordering is associated with
the nucleation of rare but very ``hot" sites in the system's volume. The
statistics of these hot sites is shown in
Fig.~\ref{fig::scaling_of_Nhot_vs_gamma}
for quenches of different strength $\kappa$, for different time moments
$t$. The data demonstrates that the hot sites are indeed rare for all
$\kappa$ and $t$. Indeed, for the system with
$\sim 10^5$
of sites, the number of hot sites never exceeds
$4\times 10^3$.

\begin{figure}[b!]
    \centering
	\includegraphics[width=0.85\columnwidth]{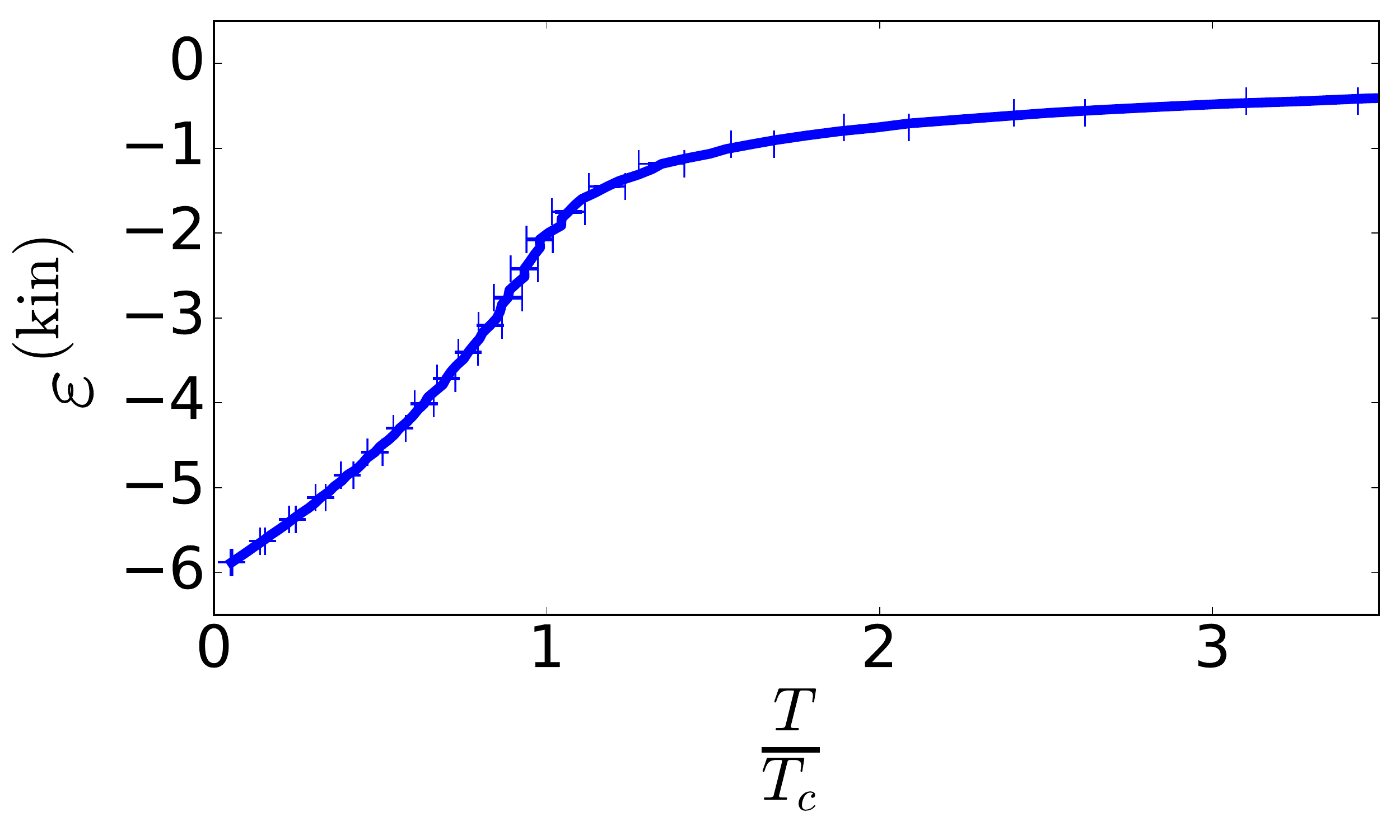}
        \caption{
\label{fig::SM_Ekin_vs_T_eq}
Equilibrium kinetic energy density 
$\varepsilon^{\rm (kin)}$
as a function of the reduced temperature $T/T_c$. The simulation is
performed for 
$g=10$, $V=50\times50\times50$.
We see that there is a one-to-one correspondence between the equilibrium
value of
$\varepsilon^{\rm (kin)}$
and $T$. This allows us to use
$\varepsilon^{\rm (kin)}$
in the main text as a measure of the temperature.
}
\end{figure}

\begin{figure}[t!]
    \centering
    \begin{tikzpicture}
    \node[inner sep=0pt] (duck) at (0,0)
    {\includegraphics[width=0.85\columnwidth]{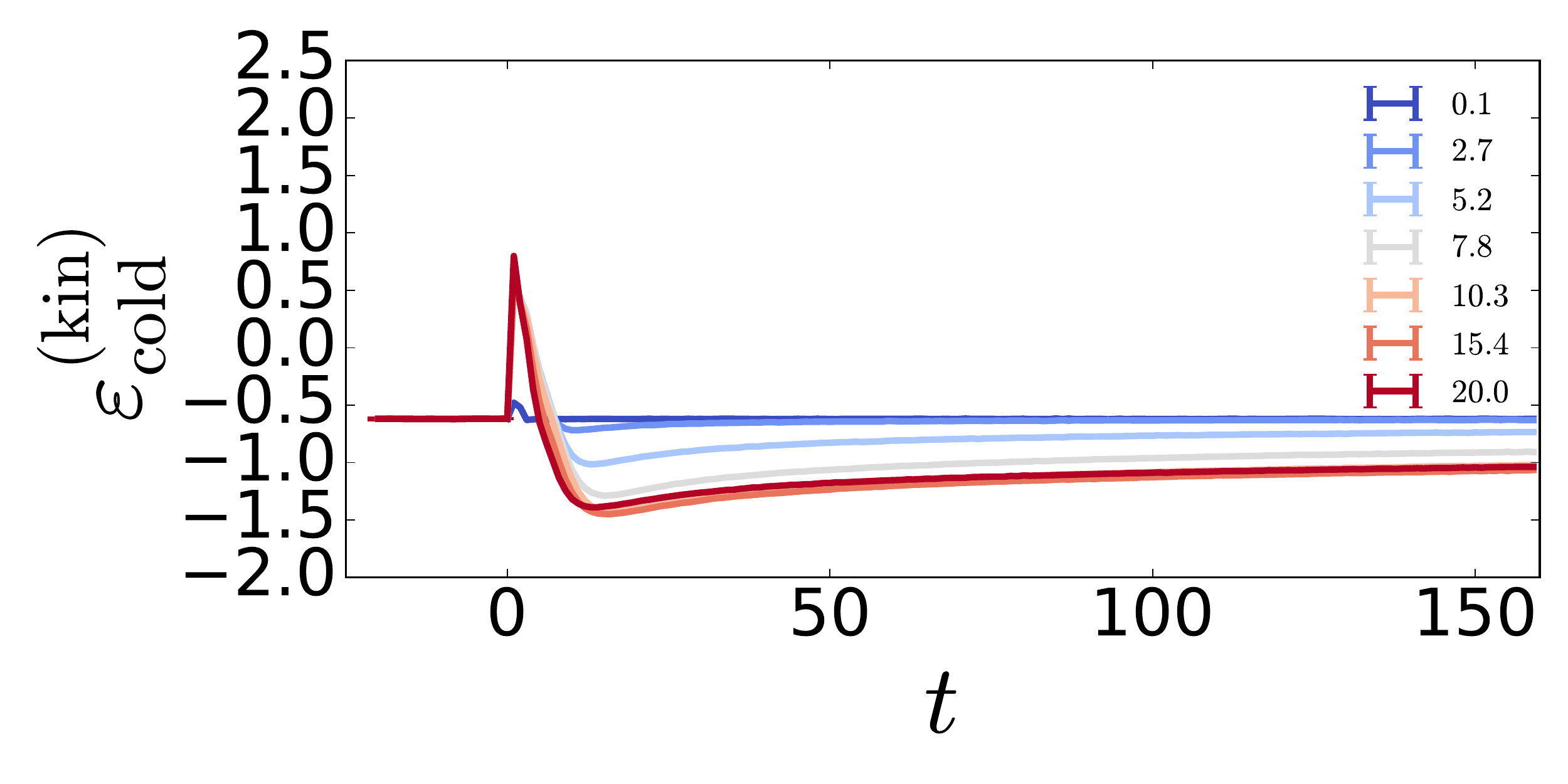}};
    \node[align=center,fill=white] at (-3.8, 1.3) {\textbf{a}};
    \end{tikzpicture}
    
    \begin{tikzpicture}
    \node[inner sep=0pt] (duck) at (0,0)
    {\includegraphics[width=0.85\columnwidth]{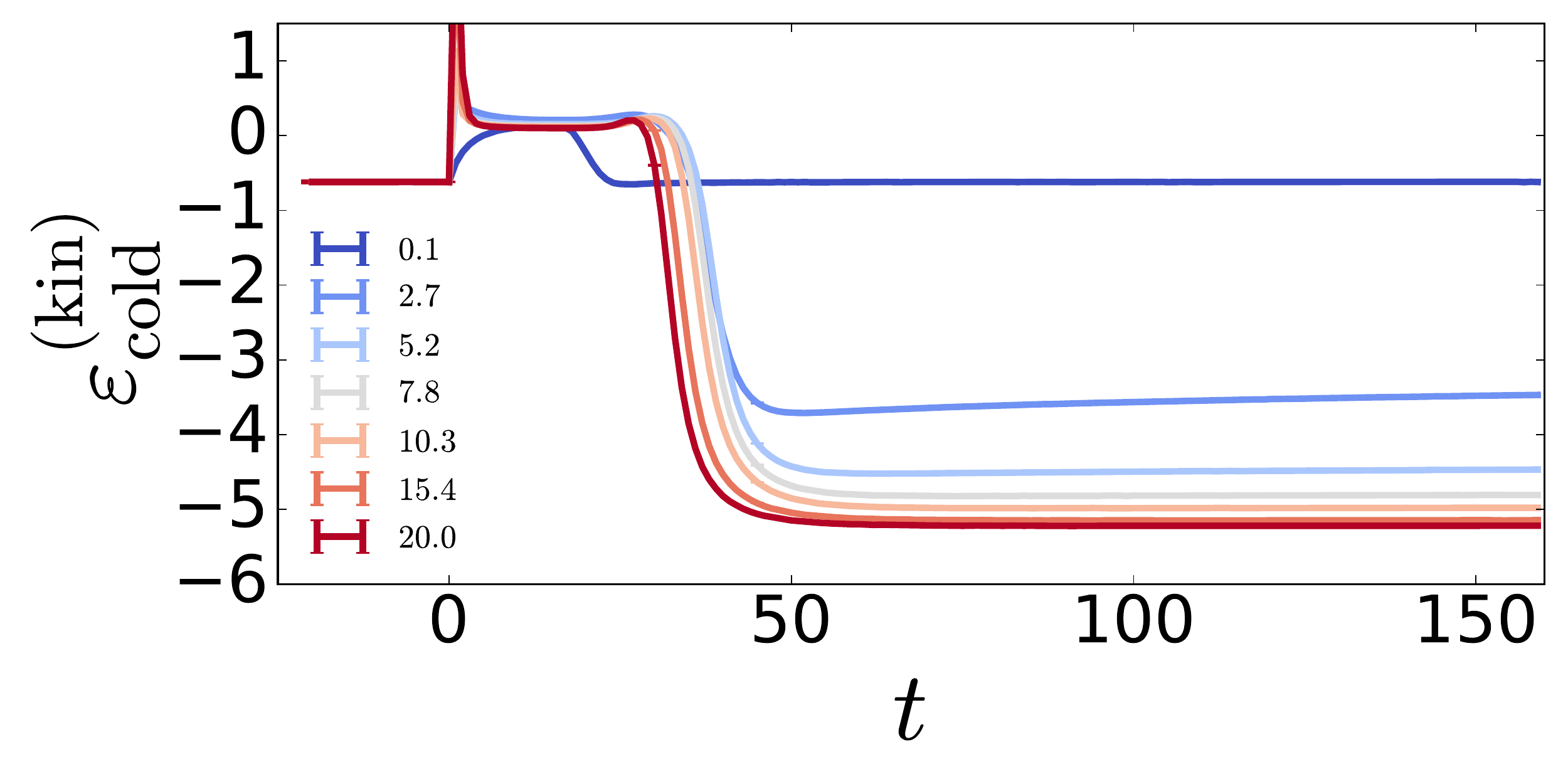}};
    \node[align=center,fill=white] at (-3.8, 1.3) {\textbf{b}};
    \end{tikzpicture}
    
    \caption{
\label{fig::Ekin_supplement}
(a) and (b) The kinetic energy $\varepsilon_{\rm cold}^{\rm (kin)}$  of the cold sites for the shorter quenches (a) and the longer quenches (b), which were  used for calculating $T_{\rm eff}(t)$ in Figs.~\ref{fig::evolution_energy_OP} (a2) and (b2) respectively. 
}
\end{figure}

As shown in the main text, the post-quench relaxation is very slow. This 
is a consequence of the hot sites being effectively decoupled from the rest
of the system. This feature can be easily understood with the help of the
following heuristic argument. When site $j$ accumulates large norm (that 
is, becomes hot) the dynamics of
$\psi_j$
may be effectively described by the equation
\begin{eqnarray}
\label{eq::hot_dgpe}
i \frac{d\psi_j}{d t} \approx g |\psi_j|^2 \psi_j.
\end{eqnarray}
This expression can be obtained from
Eq.~(\ref{eq::effect_dgpe})
in which the kinetic energy is omitted due to its smallness relative
to the nonlinear term. The solution to
Eq.~(\ref{eq::hot_dgpe})
reads
$\psi_j (t) = \sqrt{n_j} e^{-i g n_j t}$,
where the local norm
$n_j = |\psi_j|^2$
is conserved by Eq.~(\ref{eq::hot_dgpe}). We see that for large $n_j$ the local 
complex phase changes very quickly, leading to effective decoupling
of $j$ site from its neighbours.

This simple argument is confirmed by the simulations. As a numerically
computable measure of coupling between sites $i$ and $j$ we introduce the
angle
$(\psi_i, \psi_j) = {\rm arg}\, (\psi_i^* \psi_j^{\vphantom{*}})$
and estimate averaged 
$\cos^2 (\psi_i, \psi_j)$.
For totally uncorrelated (``decoupled")
$\psi_i$
and
$\psi_j$,
the averaged value is
$\langle \cos^2 (\psi_i, \psi_j) \rangle = 1/2$.
For positively correlated complex phases of
$\psi_i$
and
$\psi_j$
the averaged cosine squared exceeds 1/2, which indicates some degree of
``coupling" between $i$ and $j$.

\begin{figure}[ht]
    \centering
    \begin{tikzpicture}
    \node[inner sep=0pt] (duck) at (0,0)
    {\includegraphics[width=0.95\columnwidth]{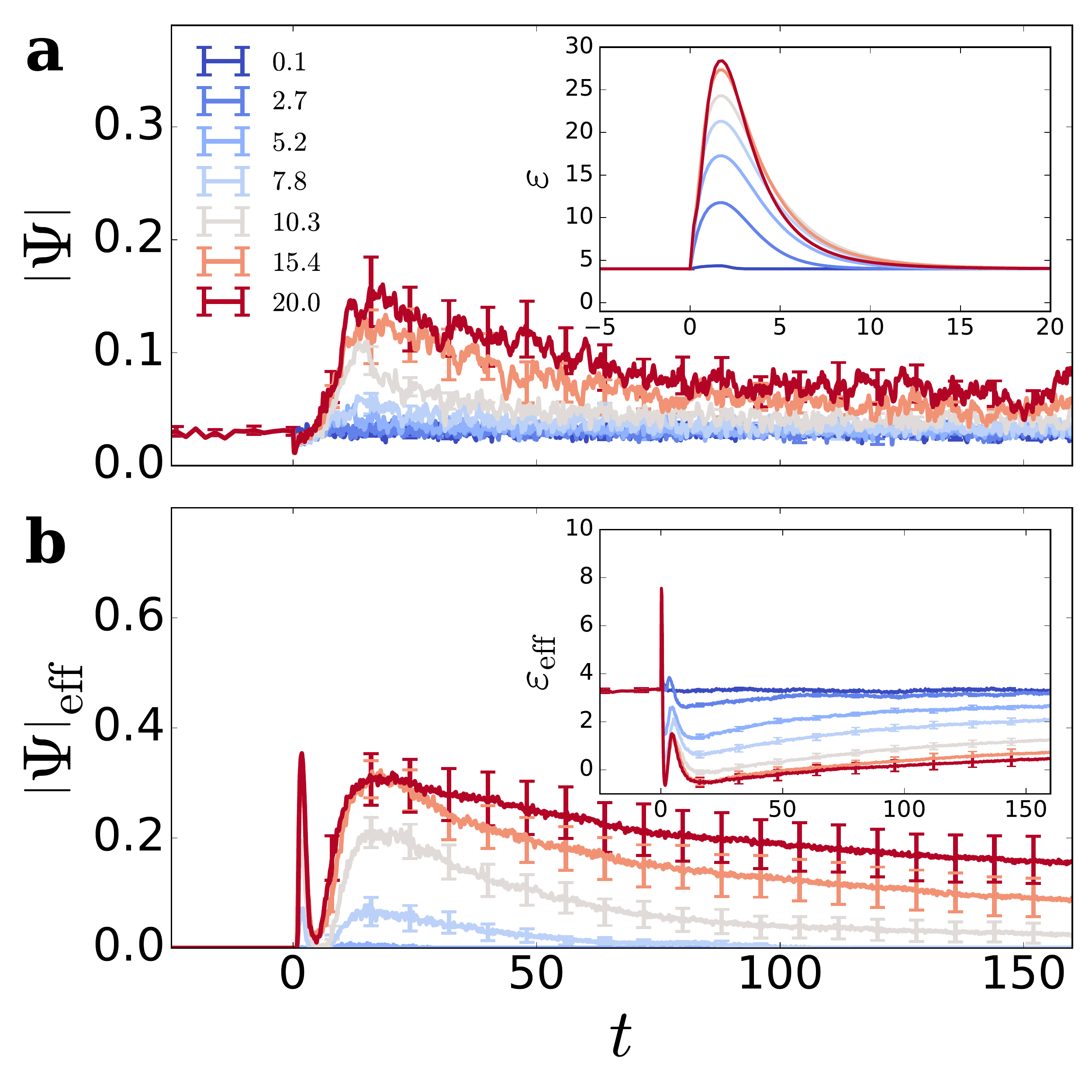}};
    \end{tikzpicture}

        \caption{
\label{fig::SM_evolution_energy_OP}
Transient ordering. (a) The order parameter 
$|\Psi|$,
barely discernible before the quench, is revived by quench (the energy
quench plotted in the inset), and demonstrates long-lasting post-quench
relaxation. (b)~The equilibrium order parameter at the effective
temperature of cold sites
$|\Psi|_{\rm eff} (t)$
defined by
Eq.~(\ref{eq::SM_EFF_OP})
exhibits similar time evolution as
$|\Psi| (t)$
in panel~(a).
Inset: The effective energy density of cold sites
$\varepsilon_{\rm eff}$.
The simulations are performed for 
$g=10$, $V=10\times10\times10$,
pre-quench temperature
$T_0/T_{\rm c} \approx 16$, the quench intensity $\kappa$ plotted in the legend.
}
\end{figure}

With this in mind, let us examine the data for 
$\cos^2 (\psi_i, \psi_j)$
plotted in
Fig.~\ref{fig::cos_correlation}.
Here $i$ and $j$ are always the nearest neighbours. When $i$ belongs to the
cold sites subset, the corresponding curves are in panel~(a), while for $i$
being hot, panel~(b) should be consulted. The presented graphs demonstrate
that, after the quench, the correlations between the neighbors for both the
cold and hot sites transiently enhance. However, the cold sites remain
correlated with their neighbors for a long time after the quench, whereas
the hot sites quickly decouple from their environment and become
uncorrelated, see
Fig.~\ref{fig::cos_correlation}\,(a) and~(b)
respectively. The large fluctuations before the quench in
Fig.~\ref{fig::cos_correlation}\,(b)
are caused by averaging over a small number of hot sites present in
equilibrium.

\begin{figure}[hb]
    \centering
	\includegraphics[width=0.75\columnwidth]{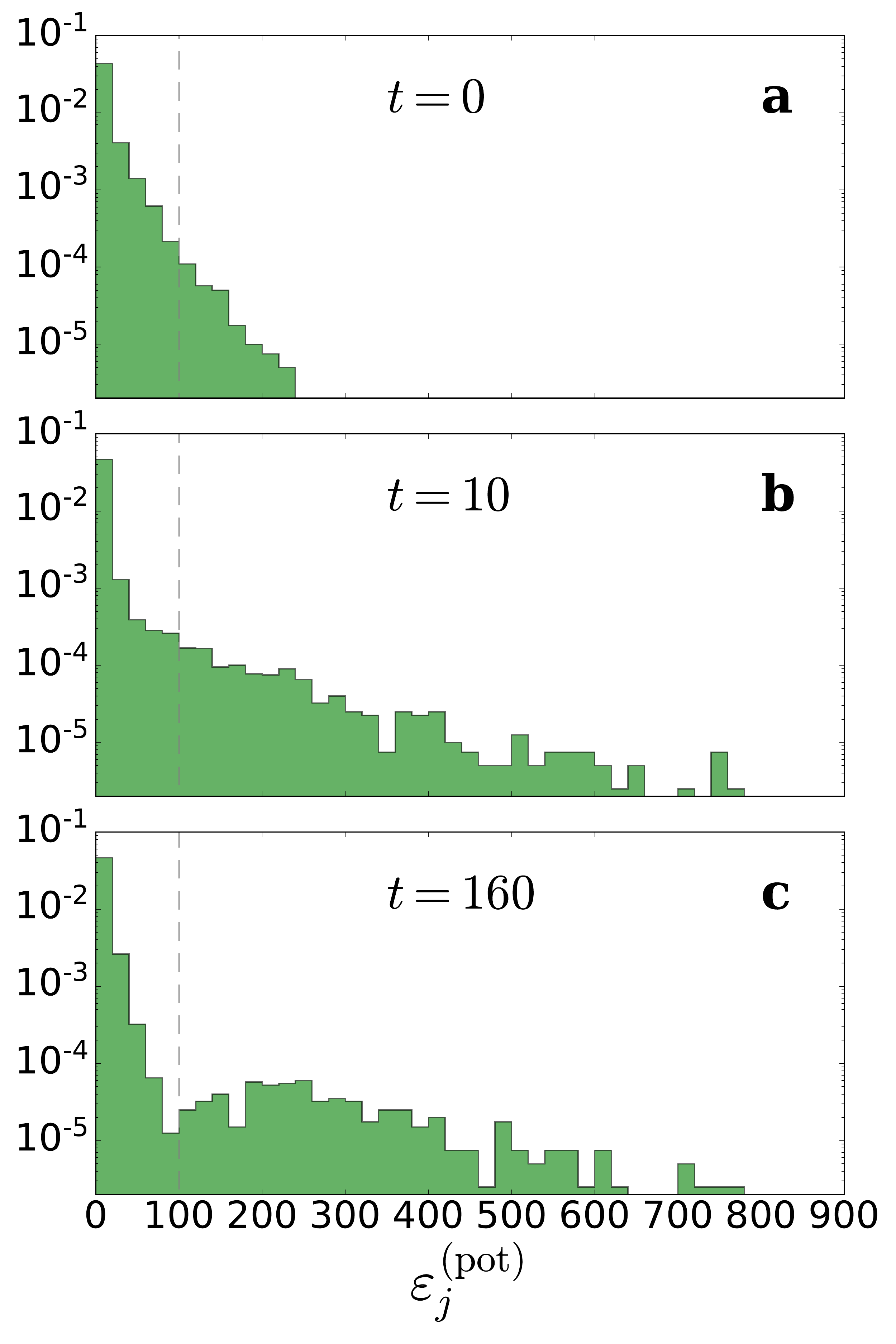}
        \caption{
\label{fig::SM_pot_energy_distr}
Snapshots of the spatial distribution of local potential energy
$\varepsilon_j^{\rm (pot)}=\frac{g}{2} |\psi_j|^4$.
(a)~Equilibrium distribution.
(b)~Distribution immediately after the quench ($t=10$).
(c)~Distribution during the post-quench relaxation ($t=160$).
The horizontal axis represents local potential energy, the vertical axis
shows the concentration of sites with a given value of
$\varepsilon_j^{\rm (pot)}$.
Presence of a larger number of sites with atypically large potential energy
is visible in (b,c). The threshold energy
$\varepsilon_{th}=100$
is marked by vertical dashed lines. The simulation is performed for 
$g=10$, $V=10\times10\times10$,
pre-quench temperature
$T_0/T_{\rm c} \approx 16$,
quench intensity
$\kappa = 20$.
}
\end{figure}

\section{Equilibrium kinetic energy density as a measure of the system's
temperature}

In the main text, in order to characterize the effective temperature of the
cold sites, we used the kinetic energy density of the cold sites. The
ability of the kinetic energy density to act as ``a thermometer" is
illustrated by the plot in
Fig.~\ref{fig::SM_Ekin_vs_T_eq}.
The plot presents the equilibrium kinetic energy density versus the
temperature calculated numerically. We see that the energy density grows
continuously and monotonically with $T$, and there is a one-to-one
correspondence between
$\varepsilon^{\rm (kin)}$
and $T$ which can be formally expressed either as
$\varepsilon^{\rm (kin)} = \varepsilon^{\rm (kin)}(T)$
or as
$T = T(\varepsilon^{\rm (kin)})$.
This relation allows one to extract $T$ by measuring 
$\varepsilon^{\rm (kin)}$,
as we did in the main text. The dynamics of $\varepsilon_{\rm cold}^{\rm (kin)}$ for the short and long quenches in Fig.~\ref{fig::evolution_energy_OP}(a2,b2) are shown in Fig.~\ref{fig::Ekin_supplement}.

\section{Transient ordering in a small cluster}

\begin{figure}[ht]
    \centering
    \begin{tikzpicture}
    \node[inner sep=0pt] (duck) at (0,0)
    {\includegraphics[width=0.85\columnwidth]{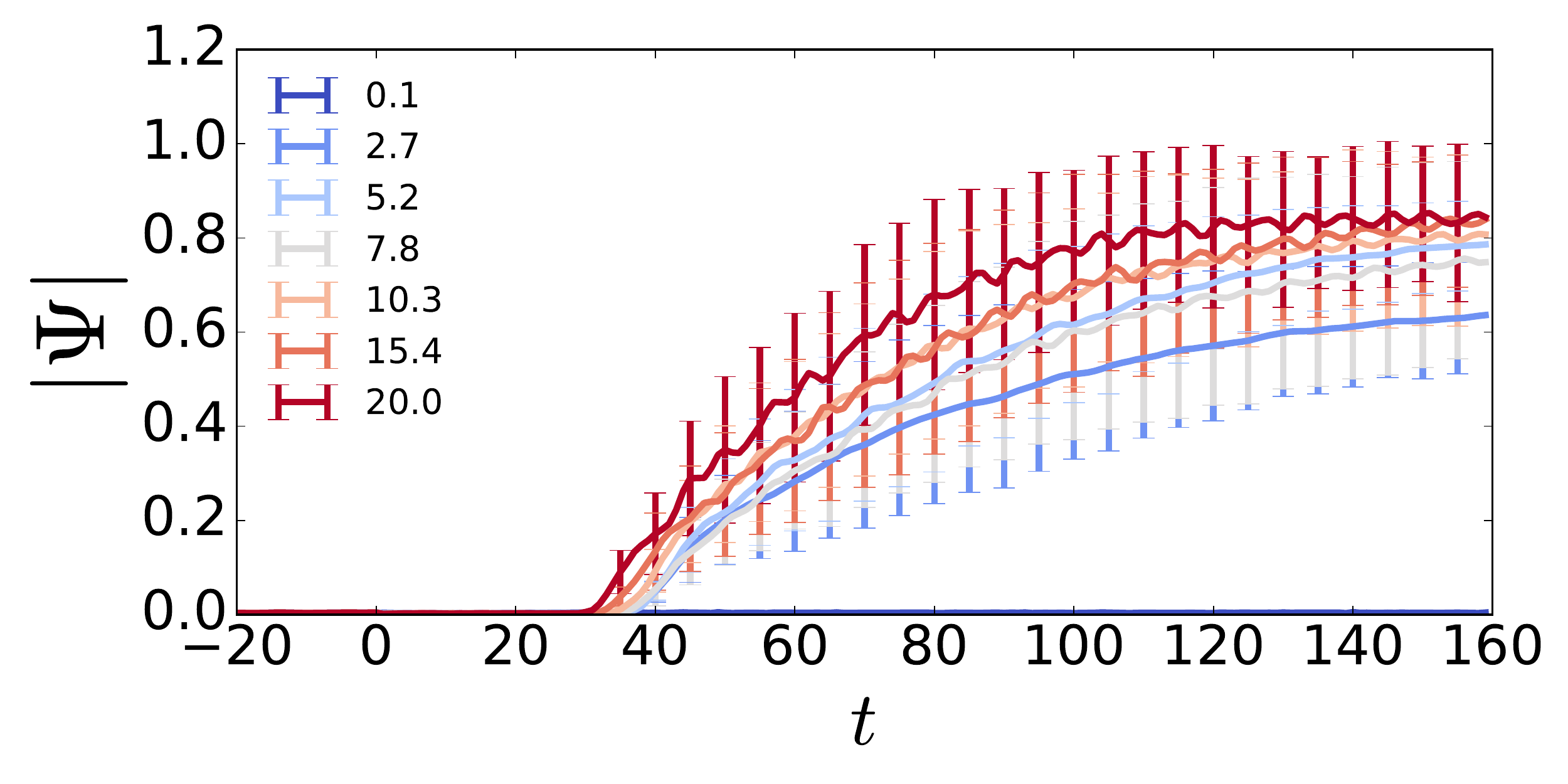}};
    \node[align=center,fill=white] at (-3.8, 1.3) {\textbf{a}};
    \end{tikzpicture}
    
    \begin{tikzpicture}
    \node[inner sep=0pt] (duck) at (0,0)
    {\includegraphics[width=0.85\columnwidth]{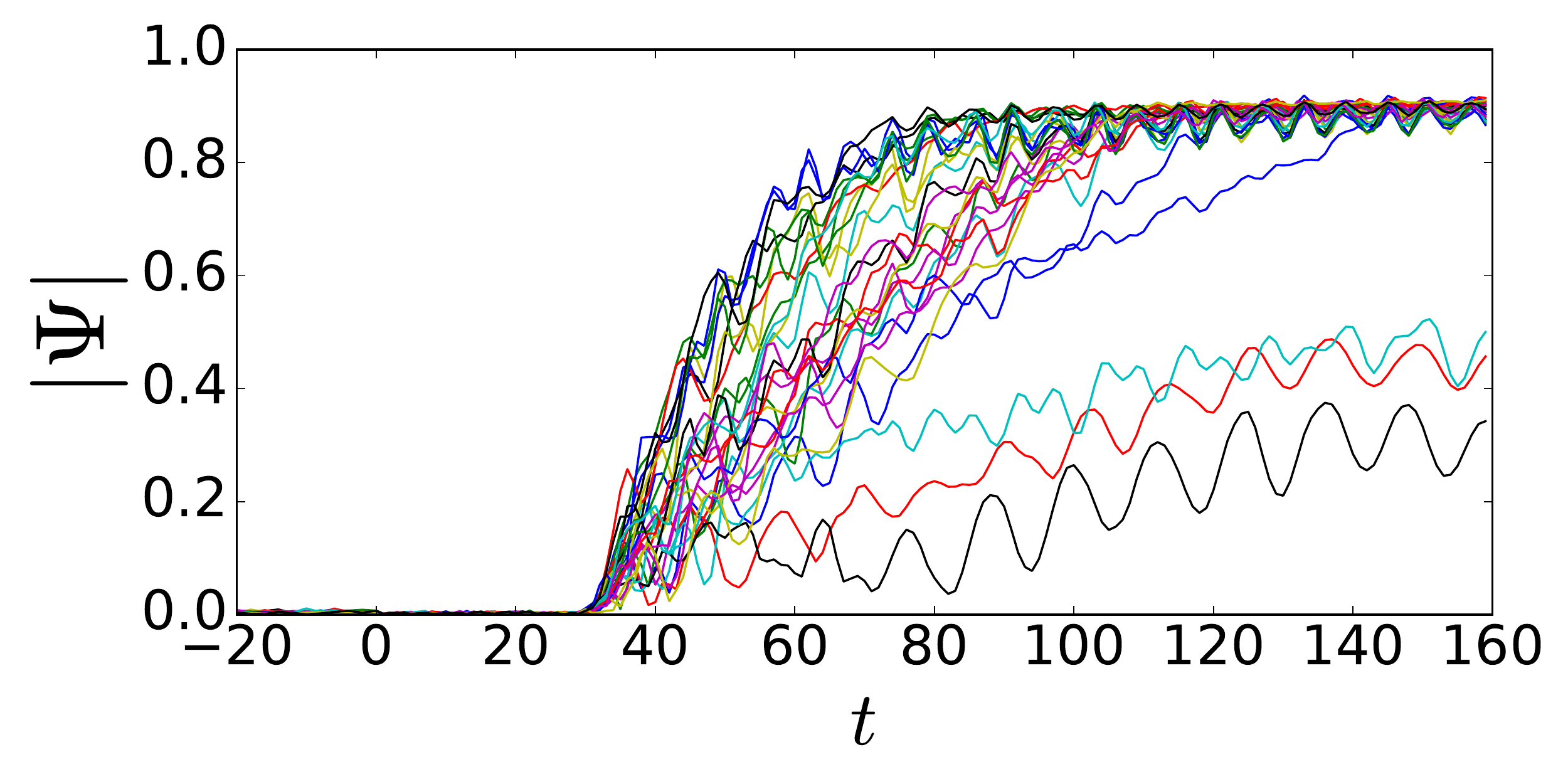}};
    \node[align=center,fill=white] at (-3.8, 1.3) {\textbf{b}};
    \end{tikzpicture}

        \caption{
\label{fig::Large_system_long_quench}
Transient ordering with the emergence of $U(1)$ order parameter after the long quench from Fig.~\ref{fig::evolution_energy_OP}(b1,b2). (a) The order parameter 
$|\Psi|$, barely discernible before the quench, is revived by the quench. (b)~Realizations of $|\Psi|$ for different initial conditions for $\kappa=20$ from (a). The simulations are performed for the pre-quench temperature
$T_0/T_{\rm c} \approx 2.3$, the quench intensity $\kappa$ plotted in the legend.}
\end{figure}

The transient ordering in small clusters requires special attention due to $l_c$ being larger than the cluster
linear size. In such a situation, it is more convenient to follow the 
evolution of the order parameter
$|\Psi| (t)$,
rather than
$l_c$.
For a small system the order-parameter response to the quench is
illustrated by
Fig.~\ref{fig::SM_evolution_energy_OP}(a).
It shows
$|\Psi|(t)$
and
$\varepsilon(t)$
(in the inset) for
$g=10$
and
$T_0/T_{\rm c} \approx 16$.
When
$t<0$,
the equilibrium value of $|\Psi|$ is very small, determined by finite-size
effects. Once the quench is launched at
$t=0$, the system energy spikes, the order parameter initially drops, but quickly
starts growing. It reaches the maximum value ($|\Psi| \approx 0.15$ for the strongest quench) at $t \sim 20$. Remarkably, the post-quench relaxation of
$|\Psi|$ to its equilibrium value is very slow.

In
Fig.~\ref{fig::SM_pot_energy_distr},
the snapshots of the potential energy distribution $\varepsilon_j^{\rm (pot)}$ in equilibrium and at two different time
moments are presented
($g=10$, $V=10\times10\times10$, $T_0/T_{\rm c} \approx 16$, $\kappa = 20$). If one sets
$\varepsilon_{th} = 100$,
the equilibrium state at
$T_0/T_{\rm c} \approx 16$,
contains
$x_{\rm hot}^{\rm eq} \approx 0.55\%$
of hot sites, the energy accumulated there is
$E_{\rm hot}^{\rm eq} \approx 685$,
which is approximately
$17\%$
of the total energy
$E=4000$.
The quench acts to increase these values. For the state
in
Fig.~\ref{fig::SM_pot_energy_distr}(b),
one has
$x_{\rm hot} \approx 1.3 \%$
and
$E_{\rm hot} \approx3500$,
which is approximately
$88\%$
of the total energy
$E=4000$.

To characterize the transient states of the small cluster, we use the effective energy
density
$\varepsilon_{\rm eff} (t)
=
[E(t) - E_{\rm hot}(t)]/ \left(1 - x_{\rm hot}\right) V$,
plotted in the inset of
Fig.~\ref{fig::SM_evolution_energy_OP}(b),
and the effective equilibrium temperature corresponding to the energy density of cold sites
$T(\varepsilon_{\rm eff})$. When 
$T(\varepsilon_{\rm eff}) \sim T_{\rm c}$,
the emergence of the order parameter may be expected. Indeed, if we define
``effective" order parameter as follows
\begin{eqnarray}
\label{eq::SM_EFF_OP}
|\Psi|_{\rm eff} = |\Psi|^{\rm eq} (T (\varepsilon_{\rm eff})),
\end{eqnarray}
we discover that the time evolution of
$|\Psi|_{\rm eff} (t)$ 
is qualitatively the same as 
$|\Psi| (t)$,
see
Fig.~\ref{fig::SM_evolution_energy_OP}(b).

\section{Emergence of U(1) order for longer quenches in large systems}

For the longer quenches shown in Fig.~\ref{fig::evolution_energy_OP}(b1,b2), we observe the emergence of non-zero $U(1)$-order parameter extending over the most of the simulated lattice, which is supposed to be a finite-size effect. In a thermodynamically large system,   when $T_{\rm eff}$ for the cold part drops below $T_c$, a transient $U(1)$ order emerges spontaneously and hence incoherently in distant parts of the system, which means that it is supposed to have a finite coherence length $l_c$ limited by the presence of $U(1)$ vortices~\cite{kibble1976topology, kibble1980some, zurek1985cosmological,
zurek1996cosmological,tarkhov2022dynamics}.

In Fig.~\ref{fig::Large_system_long_quench}, we show the behavior of the order parameter for individual quench realizations. The slower growth of order parameter for some realizations is an implicit indication of the creation of vortex loops during the quench (see discussion in the related paper~\cite{tarkhov2022dynamics}).

\end{document}